\documentclass[a4paper,12pt]{article}
\usepackage{jheppub}

\usepackage{braket}
\usepackage{subfigure}
\usepackage{simplewick}
\usepackage{mathrsfs,extarrows}
\usepackage{amsmath}
\usepackage{amssymb}
\usepackage{mathtools}
\allowdisplaybreaks[4] 
\usepackage{euscript}
\usepackage{tensor}
\usepackage{amsthm}
\usepackage{bbm}
\usepackage{graphicx}
\usepackage{tikz}

\usepackage[most]{tcolorbox}
\tcbuselibrary{skins,breakable,theorems}

\usepackage{hyperref}
\hypersetup{
    colorlinks=true,
    allcolors=blue
}

\newcommand\td{\text{d}}
\newcommand{\p}{\partial}
\newcommand{\OmegaH}{\Omega_{\textrm{\tiny H}}}

\def\>{\rangle} \def\<{\langle}
\title{\boldmath Holography in flipped AdS/$\mathbb{Z}$: Another approach to dS holography}

\author[a,b,c]{Bin Chen}
\author[b]{, Zezhou Hu}
\author[b]{, Xin-Cheng Mao}
\author[b,c]{, Shan-Ming Ruan}
\affiliation[a]{Institute of Fundamental Physics and Quantum Technology, \\\&  School of Physical Science and Technology, \\ Ningbo University, Ningbo, Zhejiang 315211, China}
\affiliation[b]{School of Physics, Peking University, \\No.5 Yiheyuan Rd, Beijing 100871, P.~R.~China}
\affiliation[c]{Center for High Energy Physics, Peking University, \\No.5 Yiheyuan Rd, Beijing 100871, P.~R.~China}
\emailAdd{chenbin1@nbu.edu.cn, z.z.hu@pku.edu.cn, 
maoxc1120@stu.pku.edu.cn, ruanshanming@pku.edu.cn}

\abstract{Motivated by the subtleties in the conventional dS/CFT correspondence, we explore analytic continuation as a constructive route to de Sitter holography. We show that quantum field theories in de Sitter space and in a spacetime that we call flipped $\mathrm{AdS}/\mathbb{Z}$ (fAdS) are related by analytic continuation. We then develop a holographic description of fAdS in terms of a boundary theory referred to as flipped CFT (fCFT). In particular, we construct the extrapolate dictionary for general asymptotically fAdS spacetimes and compute the holographic two-point functions, finding agreement with an independent derivation based on the conformal symmetry of fCFT. The analytic continuation relation between fAdS and dS further suggests that fCFT may provide a starting point for an alternative holographic description of de Sitter physics. As a nontrivial check of this picture, we show that the Cardy formula of fCFT$_2$ reproduces the Bekenstein--Hawking entropies of the cosmological horizons in both pure dS$_3$ and Kerr-dS$_3$.}

\begin{document}
\maketitle
\flushbottom

\section{Introduction and motivation}\label{sec:Intro}

The discovery that black holes and cosmological horizons carry a macroscopic entropy proportional to their area \cite{Bekenstein:1973ur,Hawking:1975iha,Gibbons:1977mu}, rather than their volume, provides one of the primary motivations for the holographic principle \cite{tHooft:1993dmi,Susskind:1994vu}, according to which a gravitational theory in the bulk may admit an equivalent description in terms of a lower-dimensional theory. From this perspective, the entropy associated with a gravitational horizon should admit a microscopic interpretation in terms of degrees of freedom in the corresponding boundary theory. This picture is realized most explicitly in the AdS/CFT correspondence \cite{Maldacena:1997re,Gubser:1998bc,Witten:1998qj}, particularly in AdS$_3$/CFT$_2$ \cite{Strominger:1997eq,Maldacena:1998bw,Carlip:1998qw}, where the Cardy formula \cite{Cardy:1986ie} for the asymptotic density of states in the boundary CFT reproduces the Bekenstein--Hawking entropy of the BTZ black hole. Related discussions of the horizon area law and its connection to the Cardy formula can be found, e.g., in Refs.~\cite{Carlip:1998wz,Carlip:1999cy}. Similar ideas also play an important role in the Kerr/CFT correspondence \cite{Guica:2008mu,Hartman:2008pb,Haco:2018ske}.

In parallel with the development of AdS/CFT, and building on earlier investigations of quantum gravity and holography in de Sitter spacetime \cite{Antoniadis:1997fu,Hull:1998vg,Maldacena:1998ih,Kim:1998zs,Banados:1998tb,Lin:1999gf,Bousso:1999cb,Bousso:2000md,Banks:2000fe,Bousso:2000nf,Balasubramanian:2001rb,Witten:2001kn}, Strominger proposed the dS/CFT correspondence \cite{Strominger:2001pn,Spradlin:2001pw}, according to which quantum gravity in de Sitter (dS) spacetime is related to a Euclidean conformal field theory defined at the spacelike conformal boundaries $\mathcal{I}^{\pm}$. One particularly suggestive piece of evidence comes from three-dimensional de Sitter gravity. It was observed in several early studies \cite{Maldacena:1998ih,Bousso:2001mw,Park:1998qk,Myung:2001ab,Kabat:2002hj} that the entropy of pure dS$_3$, namely 
\begin{equation}
	S_{\mathrm{dS}} = \frac{A_{\mathrm{hor}}}{4 G} = \frac{\pi l_{\mathrm{dS}}}{2G} \,, 
\end{equation}
can be reproduced by the Cardy formula of a putative boundary CFT$_2$. For equal left- and right-moving sectors, the thermal Cardy formula takes the form
\begin{equation}
	S=\frac{2\pi^2}{3}c \,T \,.
\end{equation}
Taking the real central charge $c=\frac{3l}{2G}$ \cite{Strominger:2001pn,Nojiri:2001mf,Klemm:2001ea,CarneirodaCunha:2001nw}, in close analogy with the Brown--Henneaux central charge in AdS$_3$/CFT$_2$ \cite{Brown:1986nw,Henningson:1998gx,Balasubramanian:1999re}, together with the dimensionless Gibbons--Hawking temperature $T=\frac{1}{2\pi}$, one indeed recovers the Bekenstein--Hawking entropy of the cosmological horizon.

Despite this appealing agreement, the interpretation of the boundary theory in dS/CFT is considerably more subtle than in its AdS counterpart. In alternative formulations of the correspondence, the analytic continuation from AdS suggests an imaginary central charge,
\begin{equation}
	c=i\frac{3l_{\mathrm{dS}}}{2G} \,,
\end{equation}
as discussed, for example, in Refs.~\cite{Balasubramanian:2001nb,Balasubramanian:2002zh}. If one simultaneously keeps the Gibbons--Hawking temperature real, a direct application of the conventional thermal Cardy formula would then produce an imaginary entropy. The successful reproduction of the dS entropy by the Cardy formula therefore does not by itself provide an unambiguous microscopic interpretation in terms of an ordinary unitary Euclidean CFT.

A further subtlety arises at the level of the holographic dictionary. In AdS/CFT, the GKPW prescription \cite{Gubser:1998bc,Witten:1998qj}, which identifies boundary correlation functions through variations of the bulk partition function, is equivalent to the extrapolate dictionary (BDHM dictionary) \cite{Banks:1998dd}, which relates boundary operators directly to the asymptotic behavior of bulk fields. In dS/CFT, however, the corresponding prescriptions are generally inequivalent \cite{Harlow:2011ke}. This distinction indicates that the relation between bulk observables and boundary quantities in de Sitter space is structurally different from the familiar AdS/CFT correspondence.

These observations suggest that, although a Euclidean CFT associated with $\mathcal{I}^{\pm}$ captures important aspects of de Sitter physics, its interpretation as a conventional holographic dual remains subtle. It is therefore natural to ask whether dS bulk physics may admit a different holographic description, in which the lower-dimensional field theory and the corresponding bulk--boundary dictionary emerge through an alternative construction. This question provides the main motivation for the approach developed in this work.

The subtleties of the conventional dS/CFT correspondence discussed above motivate us to explore alternative routes to de Sitter holography. The central idea of our work is to use analytic continuation as a constructive principle for holography. The motivation comes from the standard relation between AdS/CFT and dS/CFT, where an analytic continuation in the bulk induces a corresponding continuation of the holographic description on the boundary CFT side.  We therefore propose the following general principle: if two bulk theories are related by a controlled analytic continuation and one of them admits a holographic dual, then the same bulk continuation may be used to induce a corresponding map between the boundary theories, thereby constructing a holographic description of the second bulk spacetime. Schematically, one may express this principle as
\begin{equation}
	\boxed{
		\begin{array}{ccc}
			\mathcal{M}_{a}
			&
			\xrightarrow{\qquad \mathcal{A}_{\mathrm{bulk}} \qquad}
			&
			\mathcal{M}_{b}
			\\[10pt]
			\Big\updownarrow\mathrlap{\;\mathcal{H}_{a}}
			&&
			\mathllap{\mathcal{H}_{b}\;}\Big\updownarrow
			\\[10pt]
			\mathrm{QFT}_{a}
			&
			\xrightarrow{\qquad \mathcal{A}_{\mathrm{bdry}} \qquad}
			&
			\mathrm{QFT}_{b}
		\end{array}
	}
	\label{eq:general-holographic-continuation}
\end{equation}
where $\mathcal{H}_{a,b}$ denote the corresponding holographic dictionaries. The crucial requirement is that $\mathcal{A}_{\mathrm{bdry}}$ should be induced by $\mathcal{A}_{\mathrm{bulk}}$ through the holographic dictionaries. In this sense, a known holographic description of $\mathcal{M}_{a}$ may provide a constructive starting point for obtaining a holographic description of $\mathcal{M}_{b}$.

A natural realization of this structure is provided by the familiar relation between AdS/CFT and dS/CFT. To make the corresponding analytic continuation explicit, let us consider Euclidean AdS and de Sitter spacetime in the Poincar\'e patch, where their metrics take the form
\begin{equation}
	\td s_{\mathrm{EAdS}}^2	=\frac{l_{\mathrm{AdS}}^2}{z^2}	\left( \td z^2+ \td\vec{x}^{\,2}	\right)\,,
	\qquad
	\td s_{\mathrm{dS}}^2	=	\frac{l_{\mathrm{dS}}^2}{\eta^2}	\left(	-\td\eta^2+\td\vec{x}^{\,2}	\right)\,,
\end{equation}
and are related by the bulk analytic continuation
\begin{equation}	\label{eq:EAdS-dS-analytic-continuation}
 \mathcal{A}_{\mathrm{bulk}} : \qquad 	l_{\mathrm{AdS}} \longrightarrow 	i l_{\mathrm{dS}}\,, \qquad z	\longrightarrow 	-i\eta \,. 
\end{equation}
The asymptotic EAdS boundary $z\rightarrow0$ is therefore naturally mapped to the late-time dS boundary $\eta\rightarrow0$, while the holographic radial direction in EAdS is analytically continued into the time direction in dS.

At the semiclassical level, the bulk continuation $\mathcal{A}_{\mathrm{bulk}}$ also induces a corresponding analytic continuation on the boundary side. Importantly, this induced map $\mathcal{A}_{\mathrm{bdry}}$ should be understood as a continuation between the boundary theories themselves, rather than merely as a transformation between their partition functions. In the present example, it takes the schematic form
\begin{equation}
		\mathcal{A}_{\mathrm{bdry}}:\qquad
		\mathrm{CFT}_{d}^{\mathrm{AdS}} \longrightarrow \mathrm{CFT}_{d}^{\mathrm{dS}}	 \,.
	\label{eq:EAdS-dS-boundary-continuation}
\end{equation}
The theory on the right-hand side is obtained by analytically continuing the data characterizing the AdS boundary CFT according to the bulk map $\mathcal{A}_{\mathrm{bulk}}$.

A simple illustration is provided by the conformal dimensions associated with a bulk scalar field. Under $l_{\mathrm{AdS}}\rightarrow i l_{\mathrm{dS}}$, one finds
\begin{equation}
	\Delta_{\pm}^{\mathrm{AdS}}
	=
	\frac{d}{2}
	\pm
	\sqrt{
		\frac{d^2}{4}
		+
		m^2l_{\mathrm{AdS}}^2
	}
	\quad\longrightarrow\quad
	\Delta_{\pm}^{\mathrm{dS}}
	=
	\frac{d}{2}
	\pm
	\sqrt{
		\frac{d^2}{4}
		-
		m^2l_{\mathrm{dS}}^2
	} .
\end{equation}
More generally, quantities characterizing the boundary theory, including the stress-tensor normalization, central charge, interaction coefficients, and correlation functions, are correspondingly analytically continued. For instance, in three bulk dimensions,
\begin{equation}
	c_{\mathrm{AdS}} =\frac{3l_{\mathrm{AdS}}}{2G_N}
	\quad\longrightarrow\quad
	c_{\mathrm{dS}}= i\,\frac{3l_{\mathrm{dS}}}{2G_N}\,,
\end{equation}
up to the conventions adopted in the dS/CFT dictionary. Thus, $\mathcal{A}_{\mathrm{bdry}}$ maps the defining data of $\mathrm{CFT}_{d}^{\mathrm{AdS}}$ to those of the corresponding $\mathrm{CFT}_{d}^{\mathrm{dS}}$.

The second component of the bulk analytic continuation, $z\rightarrow-i\eta$, has a complementary interpretation. Since $z$ is the holographic radial coordinate rather than a boundary coordinate, this continuation maps radial evolution in EAdS to late-time evolution in dS \footnote{For example, the near-boundary expansion
\begin{equation*}
	\phi(z,\vec{x}) \sim z^{d-\Delta}\phi_{(0)}(\vec{x})+z^{\Delta}A(\vec{x})+\cdots
\end{equation*}
is continued to
\begin{equation*}
	\phi(\eta,\vec{x}) \sim (-i\eta)^{d-\Delta}\phi_{(0)}(\vec{x})+ (-i\eta)^{\Delta}A(\vec{x}) +\cdots \,,
\end{equation*}
with the conformal weights simultaneously continued as above.}. Accordingly, the AdS radial source--response structure and radial Hamilton--Jacobi evolution are mapped to the corresponding late-time data and wave-function evolution in de Sitter space. With an appropriate contour prescription, regularity in the Euclidean AdS interior is likewise continued to the condition selecting the Bunch--Davies state.

Having obtained the boundary theory $\mathrm{CFT}_{d}^{\mathrm{dS}}$ through
$\mathcal{A}_{\mathrm{bdry}}$, one can then determine its holographic relation to the dS bulk by analytically continuing the AdS/CFT dictionary. For Euclidean AdS, the GKPW relation gives
\begin{equation}
	Z_{\mathrm{CFT}_{d}^{\mathrm{AdS}}}[\phi_{(0)}]
	=
	Z_{\mathrm{EAdS}}[\phi_{(0)}] \simeq e^{-I_{\mathrm{EAdS}}^{\mathrm{ren}}[\phi_{(0)}]},
\end{equation}
where $\phi_{(0)}$ denotes the asymptotic boundary data. Applying $\mathcal{A}_{\mathrm{bulk}}$ to the bulk side gives
\begin{equation}
	e^{-I_{\mathrm{EAdS}}^{\mathrm{ren}}[\phi_{(0)}]}
	\quad\longrightarrow\quad
	e^{iS_{\mathrm{dS}}^{\mathrm{ren}}[\phi_{(0)}]}
	\equiv
	\Psi_{\mathrm{dS}}[\phi_{(0)}],
\end{equation}
while applying the induced map $\mathcal{A}_{\mathrm{bdry}}$ to the boundary side gives
\begin{equation}
	Z_{\mathrm{CFT}_{d}^{\mathrm{AdS}}}[\phi_{(0)}]
	\quad\longrightarrow\quad
	Z_{\mathrm{CFT}_{d}^{\mathrm{dS}}}[\phi_{(0)}].
\end{equation}
The analytically continued holographic dictionary therefore takes the form
\begin{equation}
		Z_{\mathrm{CFT}_{d}^{\mathrm{dS}}}[\phi_{(0)}] =	\Psi_{\mathrm{dS}}[\phi_{(0)}] \,, 
\end{equation}
which is precisely the wave-function formulation of the dS/CFT correspondence.

The standard relation between AdS/CFT and dS/CFT therefore provides an explicit realization of the map structure \eqref{eq:general-holographic-continuation}: the boundary theory dual to dS bulk spacetime is obtained through the induced analytic continuation $\mathcal{A}_{\mathrm{bdry}}$, while the corresponding dS holographic dictionary is obtained by analytically continuing the AdS/CFT correspondence. It therefore suggests a broader question: if de Sitter space is related by analytic continuation to another bulk spacetime that admits its own holographic description, can this second bulk relation provide an alternative route to dS holography?

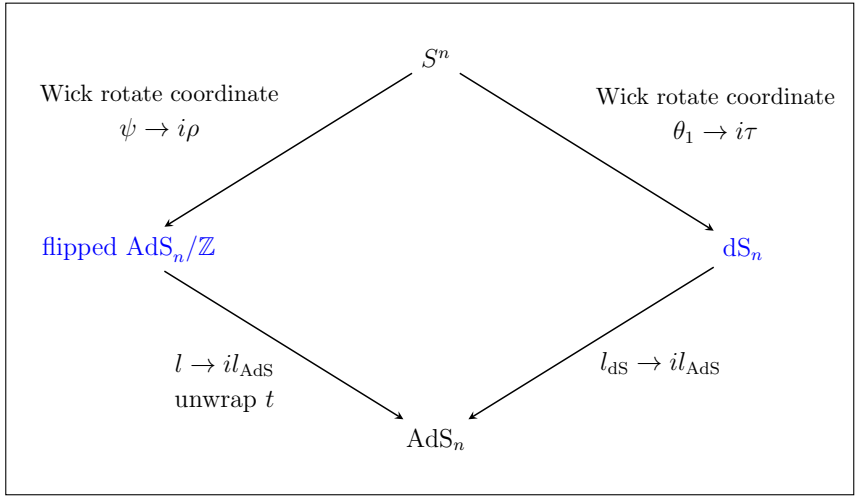
\begin{figure}[t]
	\centering
	\begin{minipage}[t]{0.75\textwidth}
	\centering
    \resizebox{\linewidth}{!}{%
    \begin{tikzpicture}[>=stealth,line width=0.7pt]
		\draw[line width=0.5pt]
		(-7,-4.0) rectangle (6.8,4.0);
		\node (S) at (0,3.1) {$S^n$};
		\node (F) at (-5.0,0) {\textcolor{blue}{$\mathrm{flipped\ AdS}_n/\mathbb{Z}$}};
		\node (dS) at (5.0,0) {\textcolor{blue}{$\mathrm{dS}_n$}};
		\node (AdS) at (0,-3.1)
		{$\mathrm{AdS}_n$};
		\draw[->](S) -- (F)
		node[midway,above left,align=center]
		{\small Wick rotate coordinate\\$\psi\rightarrow i\rho$};
		\draw[->](S) -- (dS)
		node[midway,above right,align=center]
		{\small Wick rotate coordinate\\
			$\theta_1\rightarrow i\tau$};
		\draw[->] (F) -- (AdS)
		node[midway, below left,align=center]
		{ $l \rightarrow i l_{\mathrm{AdS}}$\\ 	unwrap $t$};
		\draw[->] (dS) -- (AdS)
		node[midway,below right,align=center]{$l_{\mathrm{dS}}\rightarrow i l_{\mathrm{AdS}}$};
	\end{tikzpicture}
    }
    \end{minipage}
	\caption[Relations among sphere, de Sitter, flipped AdS, and AdS]
    {Analytic-continuation relations among $S^n$, $\mathrm{dS}_{n}$, flipped $\mathrm{AdS}_{n}/\mathbb{Z}$, and global $\mathrm{AdS}_{n}$. The sphere $S^n$ gives rise to $\mathrm{dS}_{n}$ and flipped $\mathrm{AdS}_{n}/\mathbb{Z}$ through two distinct Wick rotations. Both Lorentzian geometries are further related to $\mathrm{AdS}_{n}$ spacetime by analytic continuation of the curvature radius, together with the unwrapping of the compact time direction in the flipped-AdS construction.}
	\label{fig:geometry-continuation}
\end{figure}

Starting from the same Euclidean sphere $S^n$ with radius $l$, two distinct Wick rotations lead respectively to $\mathrm{dS}_n$ and to a Lorentzian spacetime that we refer to as flipped $\mathrm{AdS}_n/\mathbb{Z}$, abbreviated as fAdS. The resulting fAdS geometry inherits the same curvature scale $l$ from the parent sphere and is described by the metric
\begin{equation}\label{eq:flippedAdSZ}
    \td s^2/l^2=-\td \rho^2-\sinh^2\rho \td\Omega_{n-2}^2+\cosh^2\rho\td t^2\,,\quad\rho\in[0,\infty),\quad t\in[0,2\pi)\,.
\end{equation}
Locally, flipped $\mathrm{AdS}_n/\mathbb{Z}$ differs from conventional AdS$_n$ by an overall sign of the metric, while globally the coordinate $t$ is compact with period $2\pi$.\footnote{
The conventional global AdS$_n$ metric reads 
\begin{equation*}
        \td s^2/l^2_{\text{AdS}}=-\cosh^2\rho\td t^2+\td\rho^2+\sinh^2\rho\td\Omega_{n-2}^2\,,\quad\rho\in[0,\infty)\,,\quad t\in(-\infty,\infty)\,.
    \end{equation*}
Here the global time coordinate is unwrapped, as appropriate for the universal cover of AdS used in AdS/CFT. The $\mathbb{Z}$ quotient in the notation flipped $\mathrm{AdS}_n/\mathbb{Z}$ refers to the global identification $t\sim t+2\pi$.}
The analytic-continuation relations among $S^n$, $\mathrm{dS}_n$, flipped $\mathrm{AdS}_n/\mathbb{Z}$, and the conventional AdS$_n$ spacetime employed in AdS/CFT are summarized in Fig.~\ref{fig:geometry-continuation}.

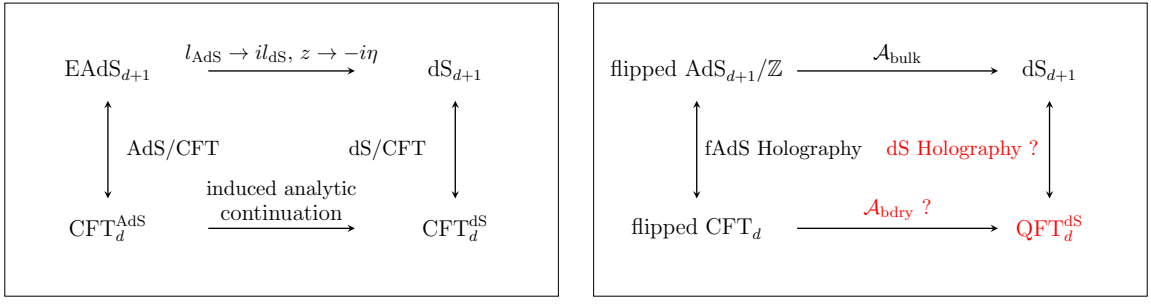
\begin{figure}[t]
    \centering
    \begin{minipage}[t]{0.485\textwidth}
        \vspace{0pt}
        \centering
        \resizebox{\linewidth}{!}{%
        \begin{tikzpicture}[>=stealth,line width=0.7pt]
            \draw[line width=0.5pt]
            (-5.1,-2.7) rectangle (5.1,2.7);
            \node (EAdS) at (-3.2,1.45)
            {$\mathrm{EAdS}_{d+1}$};
            \node (dS2) at (3.2,1.45)
            {$\mathrm{dS}_{d+1}$};
            \node (CFTAdS) at (-3.2,-1.45)
            {$\mathrm{CFT}_{d}^{\mathrm{AdS}}$};
            \node (CFTdS) at (3.2,-1.45)
            {$\mathrm{CFT}_{d}^{\mathrm{dS}}$};
            \draw[->]
            (-1.35,1.45) -- (1.35,1.45)
            node[midway,above,align=center]
            {\small $l_{\mathrm{AdS}}\rightarrow i l_{\mathrm{dS}}$, $z\rightarrow-i\eta$};
            \draw[<->]
            (-3.2,0.9) -- (-3.2,-0.9);
            \node at (-2.0,0)
            {\small AdS/CFT};
            \draw[<->]
            (3.2,0.9) -- (3.2,-0.9);
            \node at (1.95,0)
            {\small dS/CFT};
            \draw[->]
            (-1.35,-1.45) -- (1.35,-1.45)
            node[midway,above,align=center]
            {\small induced analytic\\[-1mm] continuation};
        \end{tikzpicture}%
        }
    \end{minipage}
        \hfill%
    \begin{minipage}[t]{0.485\textwidth}
        \vspace{0pt}
        \centering
        \resizebox{\linewidth}{!}{%
        \begin{tikzpicture}[>=stealth,line width=0.7pt]

            \draw[line width=0.5pt]
            (-5.1,-2.7) rectangle (5.1,2.7);
            \node (FAdS) at (-3.2,1.45)
            {$\mathrm{flipped\ AdS}_{d+1}/\mathbb{Z}$};
            \node (dS) at (3.3,1.45)
            {$\mathrm{dS}_{d+1}$};
            \node (FCFT) at (-3.2,-1.45)
            {$\mathrm{flipped\ CFT}_{d}$};
            \node (QdS) at (3.3,-1.45)
            {\textcolor{red}{$\mathrm{QFT}_{d}^{\mathrm{dS}}$}};
            \draw[->]
            (-1.35,1.45) -- (2.35,1.45)
            node[midway,above]
            {\small $\mathcal{A}_{\mathrm{bulk}}$};
            \draw[->]
            (-1.35,-1.45) -- (2.35,-1.45)
            node[midway,above]
            {\small \textcolor{red}{$\mathcal{A}_{\mathrm{bdry}}\ ?$}};
            \draw[<->]
            (-3.2,0.9) -- (-3.2,-0.9);
            \node at (-1.6,0)
            {\small fAdS Holography};
            \draw[<->]
            (3.3,0.9) -- (3.3,-0.9);
            \node at (1.7,0)
            {\small \textcolor{red}{dS Holography ?}};
        \end{tikzpicture}%
        }
    \end{minipage}%
    \caption[Analytic continuation as a route to dS holography]{
   Analytic continuation as a possible constructive principle for de Sitter holography.  (a) The conventional relation between AdS/CFT and dS/CFT. The bulk analytic continuation $\mathcal{A}_{\mathrm{bulk}}$ induces a corresponding continuation $\mathcal{A}_{\mathrm{bdry}}$ from $\mathrm{CFT}_{d}^{\mathrm{AdS}}$ to $\mathrm{CFT}_{d}^{\mathrm{dS}}$, while analytic continuation of the AdS/CFT dictionary gives the wave-function formulation of dS/CFT.
   (b) Proposed construction based on the analytic continuation between flipped $\mathrm{AdS}_{d+1}/\mathbb{Z}$ and $\mathrm{dS}_{d+1}$. Once the holographic correspondence between fAdS$_{d+1}$ and fCFT$_{d}$ is established, the bulk continuation $\mathcal{A}_{\mathrm{bulk}}$ motivates an induced boundary continuation $\mathcal{A}_{\mathrm{bdry}}$ to a candidate theory $\mathrm{QFT}_{d}^{\mathrm{dS}}$, which would provide a holographic description of dS. The quantities shown in red denote the presently open part of this construction.
    }
    \label{fig:analytic-continuation-dS-holography}
\end{figure}

Importantly, the relation between fAdS and dS is not restricted to the level of the background geometries. As we will show in detail, quantum field theories on the two Lorentzian spacetimes are also related by analytic continuation, with their correlation functions obtained from distinct analytic continuations of the same Euclidean theory on $S^n$. This therefore provides a second controlled bulk map,
\begin{equation}
    \mathcal{A}_{\mathrm{bulk}}: \qquad \mathrm{fAdS}_n \longrightarrow \mathrm{dS}_n\,,
\end{equation}
which is conceptually analogous to the EAdS/dS continuation discussed above, but arises from a different analytic continuation of the common Euclidean $S^n$ geometry.

Once this bulk relation is established, the general principle in Eq.~\eqref{eq:general-holographic-continuation} suggests a natural route toward an alternative formulation of dS holography. If fAdS admits a holographic description in terms of a boundary theory, the bulk continuation $\mathcal{A}_{\mathrm{bulk}}$ may induce a corresponding continuation $\mathcal{A}_{\mathrm{bdry}}$ on the boundary side, thereby relating the fAdS boundary theory to a candidate theory dual to de Sitter space. In this construction, the fAdS/dS continuation would play a role analogous to that of the EAdS/dS continuation in the conventional construction of dS/CFT. The standard EAdS/dS construction and the proposed fAdS/dS analogue are summarized side by side in Fig.~\ref{fig:analytic-continuation-dS-holography}.

The first necessary step toward this broader program, and the main focus of the present work, is therefore to establish the holography of flipped $\mathrm{AdS}_{d+1}/\mathbb{Z}$ itself. We show that this spacetime admits a conformal boundary and propose a holographic dual description in terms of a signature-flipped boundary theory, which we refer to as flipped $\mathrm{CFT}_{d}$ (fCFT). We develop the corresponding bulk--boundary dictionary and perform several nontrivial checks of the proposed correspondence.

In Sec.~\ref{sec:embed}, we first establish the relation between quantum field theories in dS and fAdS at the level of the path integral, showing that they arise from distinct analytic continuations of the same Euclidean theory on $S^n$. Building on our previous studies of multi-time quantum field theory \cite{Chen:2025acl,Chen:2025eeh}, we then canonically quantize the theory on constant-$\rho$ hypersurfaces in fAdS and construct the corresponding $\rho$-ordered Feynman correlator. This provides an explicit realization of the analytic continuation relating correlation functions in fAdS and dS. For completeness, the canonical quantization of quantum field theory in de Sitter space is reviewed in Appendix~\ref{sec:dSQFT}.

Having established the bulk quantum field theory in fAdS, we next turn to its holographic dual description. Since the conformal boundary has topology $ T^{n-2,1}=S^{n-2}\times S^1$ and is equipped with the signature-flipped boundary metric, the corresponding dual theory is therefore referred to as a flipped CFT (denoted by fCFT). In Sec.~\ref{sec:dSentropy}, we develop the extrapolate dictionary for a free scalar field in arbitrary-dimensional asymptotically fAdS spacetimes. We then specialize to global fAdS and compute the two-point functions of the dual fCFT holographically. As an independent consistency check, Appendix~\ref{sec:flipCFT} analyzes the fCFT directly from its conformal symmetry and shows that the resulting vacuum two-point functions agree with the holographic calculation.

These results also provide motivation for exploring fCFT as a possible starting point for a holographic description of de Sitter physics. A particularly suggestive indication arises in three dimensions, where the Cardy formula of fCFT$_2$ reproduces the Bekenstein--Hawking entropy of the dS$_3$ cosmological horizon. In Sec.~\ref{sec:kerrdS}, we demonstrate this correspondence for both pure dS$_3$ and Kerr-dS$_3$. For Kerr-dS$_3$, the modular parameter of the corresponding fCFT$_2$ can be determined by implementing the same discrete quotient on global dS$_3$ and global fAdS$_3$. An equivalent derivation based on appropriate coordinate transformations of the two global geometries is presented in Appendix~\ref{sec:conifoldS3} as an independent check. The agreement for both pure dS$_3$ and Kerr-dS$_3$ provides nontrivial evidence that fCFT$_2$ captures microscopic information associated with dS$_3$ cosmological horizons, and further motivates the broader program of constructing de Sitter holography through the induced analytic continuation described above.

\noindent\textbf{Note added:}
While this work was being completed, Ref.~\cite{Fujiki:2026kca} proposed a closely related approach to de Sitter holography based on a two-time geometry whose conformal boundary is a Lorentzian torus. In three dimensions, the corresponding bulk geometry coincides with the global flipped $\mathrm{AdS}_3/\mathbb{Z}$ spacetime considered in the present work. Ref.~\cite{Fujiki:2026kca} also presents several nontrivial checks of this holographic framework, including the reproduction of de Sitter entropy, correlation functions, timelike entanglement entropy and pseudo-entropy, and discusses a possible extension to higher dimensions. Although there is some overlap in the 3D geometry and some of the resulting observables, the two constructions are motivated from different starting points and develop de Sitter holography from different perspectives.

\section{Bulk Theory in flipped AdS/$\mathbb{Z}$ space} \label{sec:embed}

\subsection{Connections between flipped AdS/$\mathbb{Z}$ and dS space}
In this section, we first define the QFTs in dS and flipped AdS/$\mathbb{Z}$ through the path-integral approach. The Feynman correlation functions are defined separately in these two different cases. We will see that they can be related to each other through analytic continuation. The maximally symmetric Euclidean sphere $S^n$ plays an important role throughout the constructions.

The sphere $S^n$ can be embedded in the Euclidean space $\mathbb R^{n+1}$ as
\begin{equation} \label{eq:embedSn}
    \td s^2 =(\td x^1)^2+ \cdots +(\td x^{n+1})^2, \quad (x^1)^2+ \cdots +(x^{n+1})^2 = l^2\,.
\end{equation}
Similar to the treatment in flat space, we can parameterize and complexify $x^1, \cdots, x^{n+1}$ in two different ways.
\begin{figure}[htbp]
    \centering
    \includegraphics[width=1.0\linewidth]{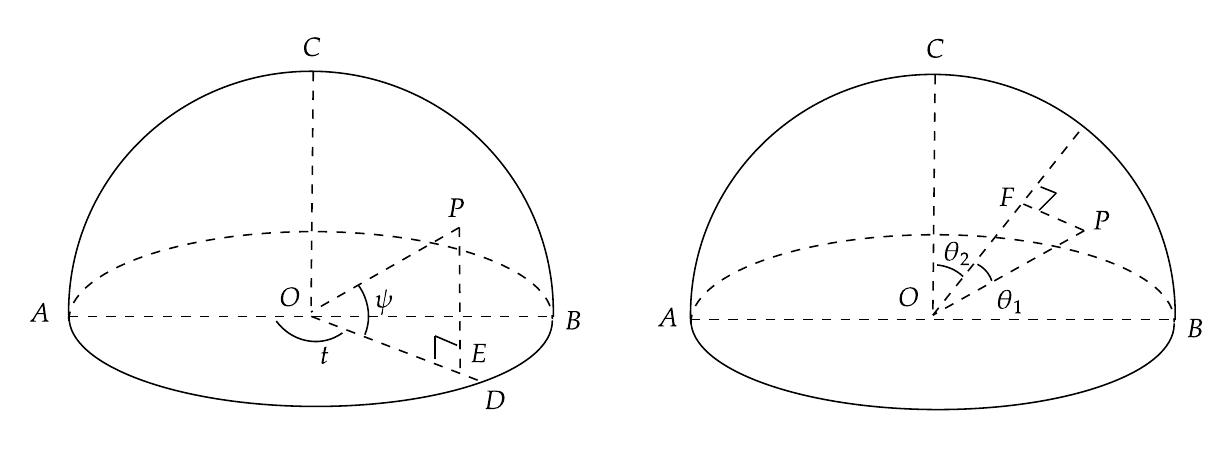}
    \caption{For a general point $P$ on the hemi-$S^2$, we project $P$ onto the plane $AOBD$, with projection point $E$. The angle $\psi$ is defined as $\angle POE$, while $t=\angle AOD$. Similarly, projecting $P$ onto the plane $AOBC$ at $F$, we define $\theta_1= \angle AOD$ and $\theta_2=\angle COF$.}
    \label{fig:hemi-S2}
\end{figure}

One way is
\begin{equation} \label{eq:Sn}
    \begin{aligned}
        &(x^i,x^n,x^{n+1})=l(y^i\cos\theta_1\cos\theta_2,\cos\theta_1\sin\theta_2,\sin\theta_1),\quad 1\leq i\leq n-1\\
        &\td s^2/l^2 = \td \Omega_n^2 =\td \theta_1^2+\cos^2\theta_1 \td\Omega_{n-1}^2 =\td \theta_1^2+\cos^2\theta_1 (\td\theta_2^2+\cos^2\theta_2\td\Omega_{n-2}^2)
    \end{aligned}
\end{equation}
where $y^i$ are the embedding coordinates of the unit $S^{n-2}$, 
\begin{equation}
    \begin{cases}
        y^1&=\sin\theta_3\\
        y^2&=\cos\theta_3\sin\theta_4\\
        \vdots\\
        y^{n-2}&=\cos\theta_3\cos\theta_4\cdots\cos\theta_{n-1}\sin\theta_n\\
        y^{n-1}&=\cos\theta_3\cos\theta_4\cdots\cos\theta_{n-1}\cos\theta_n\\
    \end{cases}\\
\end{equation}
The ranges of these angles are $\theta_i \in [-\frac{\pi}{2}, \frac{\pi}{2}] (i<n),~\theta_n \in [0, 2\pi)$.
    
The complexification $\theta_1\rightarrow i \tau$ in Eq.~\eqref{eq:Sn} gives
\begin{equation}
    \begin{aligned}
        &(x^i,x^n,x^{n+1}) =l (y^i\cos\theta_1\cos\theta_2,\cos\theta_1\sin\theta_2,\sin \theta_1)\\
        &\longrightarrow (x^i,x^n,t^{n+1})=l(y^i\cosh\tau\cos\theta_2,\cosh\tau\sin\theta_2,\sinh\tau)\,,
    \end{aligned}
\end{equation}
where $x^{n+1} = i t^{n+1}$ after the complexification. Then the embedding \eqref{eq:embedSn} becomes
\begin{equation}
    \td s^2 = (\td x^1)^2 + \cdots + (\td x^n)^2 -(\td t^{n+1})^2, \quad (x^1)^2+\cdots + (x^n)^2 -(t^{n+1})^2 = l^2\,, 
\end{equation}
which is equivalent to the definition of dS spacetime with
\begin{equation}\label{eq:metricdS}
    \td s^2/l^2=-\td \tau^2+\cosh^2\tau \td\Omega_{n-1}^2\,,\quad\tau\in(-\infty,\infty)\,.
\end{equation}
    
The other way is
\begin{equation} \label{eq:Sn2}
    \begin{split}
        & (x^i, x^n, x^{n+1}) = l(y^i \sin \psi, \cos \psi \cos t, \cos \psi \sin t), \\
        & \td s^2/l^2 = \td\Omega_n^2 =\td \psi^2+\sin^2\psi \td\Omega_{n-2}^2+\cos^2\psi\td t^2
    \end{split}
\end{equation}
where $\psi\in[0,\pi/2],~ t \in[0,2\pi)$.
This metric for $S^n$ can be related to \eqref{eq:Sn} by the transformation
\begin{equation}\label{eq:reparameterization}
    \sin\psi=\cos\theta_1\cos\theta_2\,,\quad \sin t=\frac{\sin\theta_1}{\sqrt{1-\cos^2\theta_1\cos^2\theta_2}}\,.
\end{equation}
This is a reparameterization of the two-dimensional slice of fixed $y^i$ (or $\Omega_{n-2}$), which is a hemi-$S^2$, shown in Fig.~\ref{fig:hemi-S2}.

Complexification $\psi\rightarrow i \rho$ in \eqref{eq:Sn2} gives
\begin{equation}
    \begin{aligned}
        &\qquad(x^i, x^n, x^{n+1}) =l (y^i \sin \psi,  \cos \psi \cos t,  \cos \psi \sin t)\\
        &\longrightarrow (t^i, x^n, x^{n+1}) =l (y^i \sinh \rho,  \cosh\rho \cos t,  \cosh\rho \sin t),
    \end{aligned}
\end{equation}
where $x^i = i t^i$ after the complexification. The complexification changes the embedding \eqref{eq:embedSn} as
\begin{equation}
    \begin{split}
        & \td s^2 = -(\td t^1)^2 -\cdots -(\td t^{n-1})^2 + (\td x^n)^2+(\td x^{n+1})^2, \\
        & - (t^1)^2 -\cdots - (t^{n-1})^2 + (x^n)^2+(x^{n+1})^2 = l^2,
    \end{split}
\end{equation}
or equivalently,
\begin{equation}\label{eq:metricAdSZ}
    \td s^2/l^2=-\td \rho^2-\sinh^2\rho \td\Omega_{n-2}^2+\cosh^2\rho\td t^2\,,\quad\rho\in[0,\infty),\quad t\in[0,2\pi)\,.
\end{equation}
This is the flipped AdS/$\mathbb{Z}$ spacetime we have introduced in Eq.~\eqref{eq:flippedAdSZ}.

The Feynman correlation functions defined in dS and in flipped AdS/$\mathbb{Z}$ spacetimes can all be obtained from analytic continuations of the correlation function on the sphere $S^n$. 

The $S^n$ space has a metric with Euclidean signature, meaning that the Euclidean Lagrangian is negative definite. We have well-defined correlation functions
\begin{equation}
    \<\phi(x_1)\cdots\phi(x_n)\>=\int D \phi \phi(x_1)\cdots\phi(x_n)\exp\left(\int_{S^n}\sqrt{g}\td^n x 
    \mathcal{L}(\phi(x))\right)
\end{equation}
Taking a free scalar field as an example, the Lagrangian is given by
\begin{equation}\label{eq:LanginCurve}
    \mathcal{L}=-\frac{1}{2}(\nabla\phi)^2-\frac{1}{2}m^2\phi^2\,.
\end{equation}
The free two-point function on $S^n$ can be derived by considering the Dyson-Schwinger equation
\begin{equation}
    (-\nabla_x^2+m^2)\<\phi(x)\phi(x')\>= \frac{1}{\sqrt{g}} \delta^n(x-x')
\end{equation}
Since $S^n$ is a maximally symmetric space, the Green function must be a function of a single variable $\sigma$, where $l\sigma$ is the physical geodesic distance between points $x$ and $x'$ on a unit $S^n$. When $x\rightarrow x'$, the delta function dominates, and the solution tends to be that in flat Euclidean space $\mathbb{R}^n$,
\begin{equation}
    \<\phi(x)\phi(x')\>\longrightarrow\frac{\Gamma(n/2-1)}{4\pi^{n/2}(l \sigma)^{n-2}}\quad\text{when}\quad\sigma\rightarrow0\,.
\end{equation}
Then, by requiring the solution to be regular on the entire sphere $S^n$, especially at the antipodal point $\sigma=\pi$, the solution is
\begin{equation} \label{eq:SnCorrelator}
    \begin{split}
        \<\phi(x)\phi(x')\>& =\frac{\Gamma(\delta_+)\Gamma(\delta_-)}{l^{n-2}2^n\pi^{n/2}\Gamma(n/2)} {_2F_1} \left(\delta_-,\delta_+;\frac{n}{2};\frac{1+\cos\sigma}{2}\right)\,,\\ 
        \delta_\pm & =\frac{n-1}{2}\pm M , \quad M = \sqrt{\frac{(n-1)^2}{4}-m^2 l^2}.
    \end{split}
\end{equation}
Before performing analytic continuations, the $i\epsilon$-prescription $\cos\sigma\rightarrow\cos\sigma-i\epsilon$ must be introduced to avoid the potential branch point at $\cos\sigma=1$.

We can consider the coordinate system $x=(\theta_1,\theta_2,y^i)$ with metric \eqref{eq:Sn}.
Setting $x'=(0,0,\delta^i_{n-1})$ for simplicity, the Euclidean correlation function becomes
\begin{equation}
    \<\phi(x)\phi(x')\>=\frac{\Gamma(\delta_+)\Gamma(\delta_-)}{l^{n-2}2^n\pi^{n/2}\Gamma(n/2)} {_2F_1} \left(\delta_-,\delta_+;\frac{n}{2};\frac{1+\cos\theta_1\cos\theta_2\cos\Omega_{n-2}}{2}\right)\,.
\end{equation}
Analytic continuation $\theta_1\rightarrow i\tau$ gives the Feynman correlation function in dS spacetime,
\begin{equation}\label{eq:dSfreePro}
    \<\phi(x)\phi(x')\>_{\text{dS}}=\frac{\Gamma(\delta_+)\Gamma(\delta_-)}{l^{n-2}2^n\pi^{n/2}\Gamma(n/2)} {_2F_1} \left(\delta_-,\delta_+;\frac{n}{2};\frac{1+\cosh(\tau-i\epsilon)\cos\theta_2\cos\Omega_{n-2}}{2}\right)\,,
\end{equation}
where $\cos\Omega_{n-2}\equiv \cos\theta_3\cos\theta_4\cdots\cos\theta_n$. Here the $i\epsilon$-prescription cannot be extracted from  $\cos\sigma$, since  $\cos\sigma$is invariant under time reversal  $\tau\to-\tau$; see Eq.~(2.10) of Ref.~\cite{Bousso:2001mw}.

We can also consider the coordinate system $x=(\psi,t,y^i)$ and $x'=(\psi',t',y'^i)$ with metric \eqref{eq:Sn2}.
The geodesic distance is given by
\begin{equation}
    \cos\sigma=(y\cdot y')\sin\psi\sin\psi'+\cos\psi \cos\psi'\cos (t-t')\,.
\end{equation}
Analytic continuation $\psi\rightarrow i\rho$ gives the Feynman correlation function in flipped AdS/$\mathbb{Z}$ spacetime,
\begin{multline} \label{eq:FAdS2pt}
    \< \phi(x) \phi(x')\>_{\text{flipped AdS}/\mathbb{Z}}=\frac{\Gamma(\delta_+)\Gamma(\delta_-)}{l^{n-2}2^n\pi^{n/2}\Gamma(n/2)} \\
    ~~~~~~\times {_2F_1} \left(\delta_-,\delta_+;\frac{n}{2};\frac{1-(y\cdot y')\sinh\rho\sinh\rho'+\cosh\rho \cosh\rho' \cos (t- t')}{2}-i\epsilon\right)\,,
\end{multline}
which simplifies to 
\begin{equation}\label{eq:rho'=t=0}
    \< \phi(x) \phi(x')\>_{\text{flipped AdS}/\mathbb{Z}}=\frac{\Gamma(\delta_+)\Gamma(\delta_-)}{l^{n-2}2^n\pi^{n/2}\Gamma(n/2)} \\
    \times {_2F_1} \left(\delta_-,\delta_+;\frac{n}{2};\frac{1+\cosh\rho  \cos t}{2}-i\epsilon\right)\,,
\end{equation}
when $\rho'=t'=0$. Here, the $i\epsilon$-prescription can be taken out of the expression, $\cos\sigma$, since $\rho\geq 0$.

Motivated by the relation between the QFTs in dS and fAdS, it is natural to further investigate the QFT in fAdS from the perspective of canonical quantization and compute the Feynman correlator. Since the Hartle-Hawking Euclidean path integral of fAdS is formulated on the Euclidean sphere $S^n$, we study the corresponding canonical quantization schemes on $S^n$. 
This provides a direct framework for relating correlation functions in dS$_n$ and fAdS$_n$. For completeness, we briefly review the canonical quantization of dS$_n$ in Appendix \ref{sec:dSQFT}. 

The advantage of working on $S^n$ is that the Hartle-Hawking Euclidean path integrals for both spacetimes are formulated on the same Euclidean manifold. However, not all physical information can be extracted from the $S^n$. For example, the holographic description in flipped AdS/$\mathbb{Z}$ discussed in the later sections requires asymptotic boundary conditions at $\rho\rightarrow\infty$, which cannot be implemented directly on $S^n$.

\subsection{QFT in fAdS$_n$ spacetime}

The flipped AdS/$\mathbb{Z}$ spacetime with metric \eqref{eq:metricAdSZ} is a multi-time spacetime, where the coordinate $t$ should not be regarded as ``time'' but rather as a compact spatial direction. We regard the coordinate $\rho$ as ``time'' and we evolve the QFT in flipped AdS/$\mathbb{Z}$ from $\rho=0$ to $\rho\rightarrow\infty$. Note that flipped AdS/$\mathbb{Z}$ has only one conformal boundary $\rho\rightarrow\infty$, meaning that it admits an S-vector just as in Klein space.

\begin{figure}[t]
    \centering
    \subfigure[The $\rho$-ordered $n$-point correlation function $\bra{\pi/2} \phi_n (x_n) \cdots \phi_1 (x_1) \ket{0}$ with $\rho_1 < \cdots < \rho_n$. \label{fig:pathCorrepsi}]{
    \includegraphics[width=0.4\linewidth]{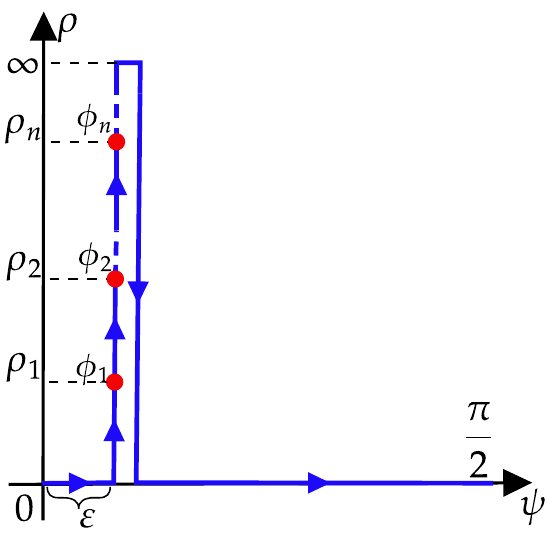}
    } \quad
    \subfigure[The $\psi$-ordered $n$-point correlation function $\bra{\frac{\pi}{2}} \phi_n (x_n) \cdots \phi_1 (x_1) \ket{0}$, where $\psi_1 < \cdots < \psi_n$. \label{fig:AdSSncorrelator}]{
    \includegraphics[width=0.4\linewidth]{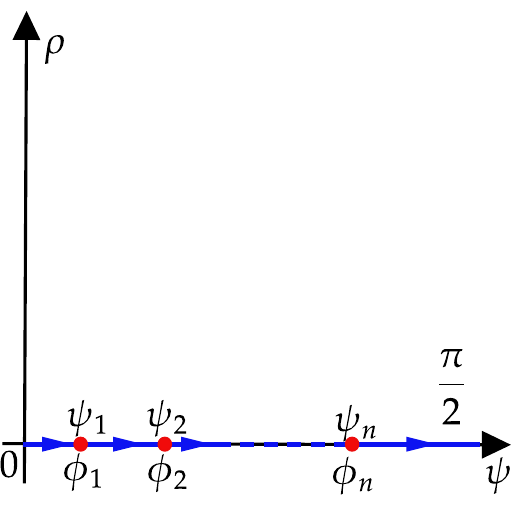}
    }
    \caption{The path-integral approach for calculating correlation functions in flipped AdS$_n/\mathbb Z$. The complex plane here is described by the complexified $\psi=i\rho$.}
\end{figure}

For the same reason as in dS spacetime, we need to focus on the Euclidean vacuum $|0\>$ associated with the slice $\rho=0$, which is given by the Euclidean path integral over the neighborhood of the slice $\psi=0$ with radius $\psi=\varepsilon$,
\begin{equation}
    \<\tilde{\phi}|0\>=\lim_{\varepsilon\rightarrow0}\int_{\phi(\psi=\varepsilon)=\tilde{\phi}} D\phi \exp\left(\int_{0}^{\varepsilon}\td\psi\int\td t\td\Omega_{n-2}\sqrt{g} \mathcal{L}(\phi)\right)\,,
\end{equation}
The Euclidean vacuum $\<\pi/2|$ associated with the Euclidean slice $\psi\rightarrow\pi/2$ is given by the path integral over the complement of the neighborhood on the sphere $S^n$,
\begin{equation}
    \<\pi/2|\tilde{\phi}\>=\lim_{\varepsilon\rightarrow0}\int_{\phi(\psi=\varepsilon)=\tilde{\phi}} D\phi \exp\left(\int_{\varepsilon}^{\frac{\pi}{2}}\td\psi\int\td t\td\Omega_{n-2}\sqrt{g} \mathcal{L}(\phi)\right)\,.
\end{equation}
The vacuum partition function is given by
\begin{equation}\label{eq:parti-AdS/Z}
    Z_{\text{flipped AdS}_n/\mathbb Z} = \<\pi/2|0\>=\int D\phi \exp\left(\int_{S^n}\sqrt{g} \td^n x \mathcal{L}(\phi)  \right)\,,
\end{equation}
The Feynman correlation functions under these vacua $\<\pi/2|P\phi(x_1)\cdots\phi(x_n)|0\>$, where $P$ is the ``time''-ordering operator placing operators with larger (smaller) $\rho$ to the left (right), are given by the path integral shown in Fig.~\ref{fig:pathCorrepsi}, where the vertical blue lines can be extended to $\rho \rightarrow \infty$. The $\rho$-ordered correlation function in flipped AdS$_n /\mathbb Z$ spacetime can also be derived from the sourced partition function
\begin{equation} \label{eq:FAdSJ}
    Z_{\text{flipped AdS}_n/\mathbb Z} [\mathcal{J}] = \int D\phi \exp\left[\int_{\mathcal{C}'} \sqrt{g} \td^n x (\mathcal{L}(\phi) + \mathcal{J} (x) \phi(x)) \right]\,,
\end{equation}
by taking functional derivatives with respect to the source $\mathcal{J} (x_i)$ with $\psi \in (\varepsilon, \varepsilon + i \infty)$. Here, $\sqrt{g} \td^n x = \sin^{n-2} \psi \cos \psi \td \psi \td^{n-2} \Omega \td t$ formally denotes the ``volume element'' of the $S^n$ metric \eqref{eq:Sn2} but with complex $\psi$, and $\mathcal{C}'$ denotes the entire blue region of complex $\psi$ in Fig.~\ref{fig:pathCorrepsi}, which cannot be deformed to the same shape as Fig.~\ref{fig:AdSSncorrelator} by Wick rotation. The reason is similar to the dS case. In particular, the vertical blue lines can be smoothly deformed to the real axis when there is no field insertion. Consequently, when $\mathcal{J} = 0$, the integral region degenerates to the entire Euclidean sphere $S^n$, and the partition function returns to \eqref{eq:parti-AdS/Z}. Comparing \eqref{eq:FAdSJ}, \eqref{eq:parti-AdS/Z} with \eqref{eq:ZdSJ}, \eqref{eq:parti-dS}, we conclude that
\begin{equation}
    Z_{\text{flipped AdS}_n/\mathbb Z} [0] = Z_{\text{dS}_n} [0], \quad Z_{\text{flipped AdS}_n/\mathbb Z} [\mathcal{J}] \neq Z_{\text{dS}_n} [\mathcal{J}].
\end{equation}

The action for a free scalar field on $S^n$ under the coordinates \eqref{eq:Sn2} can be expressed as
\begin{align*}
    S = & - \frac12 \int \td^n x \sqrt{g} \left[\nabla_\mu \phi \nabla^\mu \phi + m^2 \phi^2 \right] \\
    = & - \frac{l^{n-2}}{2} \int (\sin \psi)^{n-2} \cos \psi \td \psi \td^{n-2} \Omega \td t \left[(\p_\psi \phi)^2 + \frac{\p_a \phi \p^a \phi}{\sin^2 \psi} + \frac{(\p_t \phi)^2}{\cos^2 \psi} + m^2 l^2 \phi^2 \right],
\end{align*}
where $\td^{n-2} \Omega = \sqrt{h^{S^{n-2}}} \td^{n-2} \theta (\td^{n-2} \theta = \td \theta_3 \cdots \td \theta_n)$ and $h^{S^{n-2}}_{ab}$ are the volume element and the metric of the unit sphere $S^{n-2}$, and we used the abbreviation $x^a = (\theta_3, \cdots, \theta_n)$ to denote the coordinates on $S^{n-2}$. The conjugate momentum is then
\begin{equation}
    \pi_\psi = \frac{\delta S}{\delta (\p_\psi \phi)} = - l^{n-2} (\sin \psi)^{n-2} \cos \psi \sqrt{h^{S^{n-2}}} \p_\psi \phi.
\end{equation}
Then, the Klein-Gordon inner product is defined as
\begin{equation} \label{eq:KGfAdS}
    \begin{split}
        \< \phi^1, \phi^2 \> & = - l^{2-n} \int_{\Sigma_\psi} \td^{n-2} \theta \td t (\phi^1 \pi^2_\psi - \phi^2 \pi_\psi^1) \\
        & = (\sin \psi)^{n-2} \cos \psi \int^{2 \pi}_0 \td t \int_{S^{n-2}} \td^{n-2} \Omega \phi^1 \overset{\leftrightarrow}{\p_\psi} \phi^2
    \end{split}
\end{equation} 
The canonical quantization requires
\begin{equation} \label{eq:canonicalAdS}
    \begin{split}
        [\phi (\psi, \vec \theta, t), \pi_\psi (\psi, \vec \theta', t')] & =-\delta^{n-2} (\vec \theta - \vec \theta') \delta(t - t'), \\
        [\phi (\psi, \vec \theta, t), \phi (\psi, \vec \theta', t')] & = 0.
    \end{split}
\end{equation}

The equation of motion (Klein-Gordon equation) is
\begin{equation}
    \left[ \frac{\p_\psi ((\sin \psi)^{n-2} \cos \psi \p_\psi)}{(\sin \psi)^{n-2} \cos \psi} + (\sin \psi)^{-2} \Delta_{n-2} + (\cos \psi)^{-2} \p_t^2 - m^2 l^2 \right] \phi = 0.
\end{equation}
Here, $\Delta_{n-2}$ denotes the Laplacian operator on the unit sphere $S^{n-2}$, the eigenfunction of which is the hyperspherical function $Y^{(n-1)}_{s, \{\nu\}} (\vec \theta)$ satisfying
\begin{equation}
    \Delta_{n-2} Y^{(n-1)}_{s, \{\nu\}} (\vec \theta) = - s (s + n -3) Y^{(n-1)}_{s, \{\nu\}} (\vec \theta),
\end{equation}
where we used the abbreviation $\{\nu\} = \{\nu_1, \cdots, \nu_{n-3}\}$ and the range of the indices are $s \geq \nu_1 \geq \cdots \geq \nu_{n-3} \geq 0$. The function $Y^{(n-1)}_{s, \{\nu\}} (\vec \Omega)$ can also be chosen as real functions, which satisfies orthonormal and complete conditions similar to \eqref{eq:orthcomp}.

The field $\phi$ can be expanded through the spherical harmonics $Y^{(n-1)}_{s, \{\nu\}}$ of $S^{n-2}$ and the Fourier mode $e^{i k t}$ as
\begin{equation}
    \phi (x) = \sum_{s, \{\nu\}} \sum_{k \in \mathbb Z} Y^{(n-1)}_{s, \{\nu\}} (\vec \theta) f (\psi) e^{i k t} a_{s, \{\nu\}; k},
\end{equation}
where the constraint $k \in \mathbb Z$ is due to the periodicity $2 \pi$ of $t$, and $f (\psi)$ satisfies the following equation
\begin{equation} \label{eq:fAdS}
    -f'' + \left[ \tan \psi - \frac{n-2}{\tan \psi} \right] f' + \left[ \frac14 (n-1)^2 -M^2 + \frac{k^2}{(\cos \psi)^2} + \frac{L^2 - \frac14 (n-3)^2}{(\sin \psi)^2} \right] f = 0,
\end{equation}
with $L, M$ given by
\begin{equation} \label{eq:LM}
    L = s+\frac{n-3}{2}, \qquad M = \sqrt{\frac{(n-1)^2}{4}-m^2l^2}\,.
\end{equation}
The two solutions of this equation are
\begin{multline} \label{eq:AdSFLkM}
    F_{M, |k|, L} (\psi) = (\cos \psi)^{-L+M-1} (\sin \psi)^{L - \frac{1}{2} (n-3)} \\
    \times {_2 F_1} \left( \frac12 (1+L -|k| - M), \frac12 (1+L + |k| - M); 1+L; -(\tan \psi)^2 \right),
\end{multline}
and
\begin{multline} \label{eq:AdSGLkM}
    G_{M, |k|, L} (\psi) = \frac{\Gamma(\frac12(1+|k|+L+M)) \Gamma(\frac12(1+|k|+L-M))}{-2 (|k|!) \Gamma(1+L)} (\sin \psi)^{L-\frac{n-3}{2}} (\cos \psi)^{|k|} \\
    \times {_2F_1} \left( \frac12(1+|k|+L-M), \frac12(1+|k|+L+M); 1+|k|; \cos^2 \psi \right),
\end{multline}
which are linearly independent, since their Wronskian, defined by the $\psi$-part of the Klein-Gordon inner product \eqref{eq:KGfAdS}, is always non-vanishing
\begin{equation}
    W\left[G_{M, |k|, L}\,, F_{M, |k|, L} \right] = (\sin \psi)^{n-2} \cos \psi G_{M, |k|, L} (\psi) \overset{\leftrightarrow}{\p_\psi} F_{M, |k|, L} (\psi) =-1.
\end{equation}

Since the equation \eqref{eq:fAdS} only depends on $M^2, L^2, k^2$, we can actually obtain sixteen modes and choose two of them. Different choices are linked by Bogoliubov transformations. Note that the number $M$ in \eqref{eq:LM} can be either real or imaginary, depending on the value of the mass $m$. When $M$ is real, then $F_{M, |k|, L} (\psi),~ G_{M, |k|, L} (\psi)$ are exactly real. However, when $M$ is imaginary, $F_{M, |k|, L}^* = F_{-M, |k|, L},~ G_{M, |k|, L}^* = G_{-M, |k|, L}$. Fortunately, by using the property of the hypergeometric functions \cite{2007Table}
\begin{equation} \label{eq:2F1property}
    {_2F_1}(a,b;c;z) = {_2F_1}(b,a;c;z) = (1-z)^{c-a-b} {_2F_1}(c-a,c-b;c;z),
\end{equation}
one obtains
\begin{equation}
    F_{M, |k|, L} (\psi) = F_{-M, |k|, L} (\psi), \quad G_{M, |k|, L} (\psi) = G_{-M, |k|, L} (\psi).
\end{equation}
Therefore, only eight modes are left, and the modes are always real, regardless of whether $M$ is real or imaginary:
\begin{equation}
    F_{M, |k|, L}^* (\psi) = F_{M, |k|, L} (\psi), \quad G_{M, |k|, L}^* (\psi) = G_{M, |k|, L} (\psi).
\end{equation}

We need to pick out two appropriate modes from the eight modes $F_{M, \pm|k|, \pm L}, G_{M, \pm|k|, \pm L}$ to ensure that the regularity conditions of vacua can be expressed without any ambiguity. In this case, the two vacua $\bra{\pi/2}, \ket{0}$ are defined by Euclidean path integral by integrating over $\psi \in (\varepsilon, \pi/2), (0, \varepsilon)$ with $\varepsilon \ll 1$. The regularity conditions of the field are then
\begin{equation} \label{eq:regularAdSphi}
    \phi(\psi = 0, \vec \theta, t) \ket{0}, ~\bra{\pi/2} \phi(\psi = \pi/2, \vec \theta, t) = \text{regular}.
\end{equation}
We therefore need to select two modes that are not simultaneously divergent at both endpoints $\psi = 0, \pi/2$. It turns out that only $F_{M, |k|, L}, G_{M, |k|, L}$ with their asymptotic behavior in Table \ref{tab:fAdSmode} satisfy this criterion.
\begin{table}[htbp]
    \centering
    \begin{tabular}{c|c|c}
         \hline
         $\psi$ & $F_{M, |k|, L} (\psi)$ & $G_{M, |k|, L} (\psi)$ \\
         \hline
         0 & $0 (s\neq0)$ or $1 (s=0)$ & $\infty$ \\
         \hline
         $\pi/2$ & $\infty$ & $\frac{\Gamma(\frac12(1+L+M)) \Gamma(\frac12(1+L-M))}{-2 \Gamma(1+L)} (k=0)$ or $0 (k\neq 0)$\\
         \hline
    \end{tabular}
    \caption{The asymptotic behavior of the modes $F_{M, |k|, L} (\psi), G_{M, |k|, L} (\psi)$.}
    \label{tab:fAdSmode}
\end{table}

As a result, the mode expansion could be written as
\begin{equation} \label{eq:modeFAdS}
    \phi = \sum_{s, \{\nu\}} \sum_{k \in \mathbb Z} Y^{(n-1)}_{s, \{\nu\}} (\vec \theta) e^{i k t} \left[ F_{M, |k|, L} (\psi) a^{(F)}_{s, \{\nu\}; k} + G_{M, |k|, L} (\psi) a^{(G)}_{s, \{\nu\}; k} \right].
\end{equation}
The regularity conditions \eqref{eq:regularAdSphi} are then expressed as
\begin{equation}
    \bra{\pi/2} a^{(F)}_{s, \{\nu\}; k} = 0, \quad a^{(G)}_{s, \{\nu\}; k} \ket{0} = 0.
\end{equation}
Since $F_{M, |k|, L}, G_{M, |k|, L}$ are all real, the reality condition $\phi^* = \phi$ then yields
\begin{equation}
    a^{(F)*}_{s, \{\nu\}; k} = a^{(F)}_{s, \{\nu\}; -k}, \quad a^{(G)*}_{s, \{\nu\}; k} = a^{(G)}_{s, \{\nu\}; -k}\,.
\end{equation}

The operators $a^{(F)}, a^{(G)}$ can be re-expressed by the Klein inner product between the field and the modes:
\begin{align}
    a^{(F)}_{s, \{\nu\}; k} &= \frac{1}{2 \pi W[G_{M, |k|,L} \,, F_{M, |k|,L}]} \left\< Y^{(n-1)}_{s, \{\nu\}} (\vec \theta) e^{-i k t} G_{M, |k|,L}(\psi), \phi (\psi, \vec \theta, t) \right\> \\
    & = -(\sin \psi)^{n-2} \cos \psi \int^{2 \pi}_0 \frac{\td t}{2 \pi}\int_{S^{n-2}} \td^{n-2} \Omega Y^{(n-1)}_{s, \{\nu\}} (\vec \theta) e^{-i k t} G_{M, |k|,L}(\psi) \overset{\leftrightarrow}{\p_\psi} \phi (\psi, \vec \theta, t); \notag \\
    a^{(G)}_{s, \{\nu\}; k} & = \frac{1}{2 \pi W[F_{M, |k|,L} \,, G_{M, |k|,L}]} \left\< Y^{(n-1)}_{s, \{\nu\}} (\vec \theta) e^{-i k t} F_{M, |k|,L}(\psi), \phi (\psi, \vec \theta, t) \right\> \\
    & =  (\sin \psi)^{n-2} \cos \psi \int^{2 \pi}_0 \frac{\td t}{2 \pi}\int_{S^{n-2}} \td^{n-2} \Omega Y^{(n-1)}_{s, \{\nu\}} (\vec \theta) e^{-i k t} F_{M, |k|,L}(\psi) \overset{\leftrightarrow}{\p_\psi} \phi (\psi, \vec \theta, t). \notag
\end{align}
Then, the canonical quantization \eqref{eq:canonicalAdS} is equivalent to
\begin{equation}
    \left[ a^{(F)}_{s, \{\mu\}; k} \,, a^{(G)}_{s', \{\mu'\}; k'} \right] = \frac{l^{2-n}}{2 \pi} \delta_{s, s'} \delta_{\{\mu\}, \{\mu'\}} \delta_{k, -k'}\,.
\end{equation}

With all these data, the $\psi$-ordered two-point function can be evaluated as
\begin{multline} \label{eq:2ptAdSSn}
    \bra{\pi/2} P \phi(\psi, \vec \theta, t) \phi(\psi', \vec \theta', t') \ket{0} = - l^{2-n} \frac{\Gamma(\frac{n-3}{2})}{4 \pi^{\frac{n+1}{2}}} \bra{\pi/2} 0\> \sum_{s = 0}^\infty \sum_{k \in \mathbb Z} L C_s^{(\frac{n-3}{2})} (\vec \theta \cdot \vec \theta') \\
    \times F_{M, |k|, L} (\psi_<) G_{M, |k|, L} (\psi_>) e^{i k(t_> - t_<)},
\end{multline}
where $P$ represents the $\psi$-ordered operator that puts the field with larger $\psi$ on the left, the notation $x_<, x_>$ denote the points with smaller and larger $\psi$, respectively. Here, the Gegenbauer polynomial is defined in the same way as in \eqref{eq:Geg}. It seems that the expression \eqref{eq:2ptAdSSn} has a divergence of the form $\Gamma(\frac{n-3}{2}) \rightarrow \Gamma (0)$ when $n=3$. Fortunately, this apparent divergence is canceled by a corresponding factor of $(\Gamma(\frac{n-3}{2}))^{-1}$ that appears in the other functions in this series summation. We will see this in a special case.

Since the two-point function is SO$(n+1)$ invariant, we can always choose $(\psi',\vec\theta',t')=(0,\vec \theta',0)$ via an SO$(n+1)$ transformation. In this case, $x_> = (\psi, \vec \theta, t), ~ x_< = (0,\vec \theta',0)$, and only the $s=0$ (or $L=\frac{n-3}{2}$) mode contributes, as $F_{M, |k|, L} (\psi_< = 0) = \delta_{s, 0}$. Therefore, the free two-point function \eqref{eq:2ptAdSSn} is then rewritten as
\begin{equation}
    \begin{split}
        & \bra{\pi/2} P \phi(\psi, \vec \theta, t) \phi(\psi', \vec \theta', t') \ket{0} / \bra{\pi/2} 0\> \\
        & = l^{2-n} \frac{\Gamma(\delta_+) \Gamma(\delta_-)}{(4 \pi)^{\frac{n}{2}} \Gamma(\frac{n}{2})} {_2F_1} \left(\delta_+, \delta_-; \frac{n}{2}; \frac{1+\cos \psi \cos t}{2}\right),
    \end{split}
\end{equation}
where $\delta_\pm = \frac{n-1}{2} \pm M$, and we used $C_0^\alpha (x) \equiv 1$ and the following series expansion
\begin{multline}
    {_2F_1} \left(\delta_+, \delta_-; \frac{n}{2}; \frac{1+w\cos t}{2}\right) = \sum_{k \in \mathbb Z} \frac{2^{n-3}\Gamma(n/2)\Gamma\left( \frac12 (\delta_+ +|k|) \right)\Gamma\left(\frac{1}{2}(\delta_-+|k|) \right)}{\sqrt{\pi}(|k|!) \Gamma(\delta_+) \Gamma(\delta_-)} \\
    \times e^{ikt} w^{|k|} {_2F_1} \left( \frac12(1+|k|+L-M), \frac12(1+|k|+L+M); 1+|k|; w^2 \right).
\end{multline}
As a result, we recover Eq.~\eqref{eq:rho'=t=0} with $\psi=i\rho$ from the canonical quantization approach. Since the Feynman correlators of the dS$_n$ and flipped AdS$_n/ \mathbb Z$ spacetimes are both related to the same Euclidean correlator on $S^n$ by analytic continuation, the canonical quantization approach explicitly demonstrates the connection between the QFTs in these two spacetimes.

As a final remark, we note that AdS/$\mathbb{Z}$ spacetimes have also been discussed in Ref.~\cite{Melton:2025ecj}, where the theory is quantized by treating the coordinate $t$ as the ``time'' evolution direction. Because $t$ is periodically identified, this construction differs from the usual canonical quantization with respect to a noncompact time coordinate and is instead formulated in terms of geometric quantization. In our setup, by contrast, the metric differs by an overall sign, so that $t$ is not chosen as the evolution direction. Instead, the role of ``time'' is played by the radial coordinate $\rho$, which is noncompact and carries no periodic identification. The difficulty associated with evolution along a periodic time direction therefore does not arise in our canonical quantization of QFT in fAdS. This choice, however, comes at a different cost: evolution along $\rho$ is generically non-unitary. The QFT in fAdS should therefore be regarded as a non-Hermitian quantum system with respect to $\rho$ evolution, closely analogous to the multi-time quantum field theories in Klein space discussed in Refs.~\cite{Chen:2025acl,Chen:2025eeh}.

\section{Holographic dictionary of fAdS/fCFT correspondence}\label{sec:dSentropy}

Having formulated QFT in fAdS, we now turn to the central question of this work: whether fAdS admits a holographic description.

The metric of an asymptotically flipped AdS/$\mathbb Z$ spacetime approaches
Eq.~\eqref{eq:metricAdSZ} as $\rho \to \infty$. Its dual field theory lives on the conformal boundary $\rho\to\infty$ with metric
\begin{equation} \label{eq:T1n-2}
    \td s^2\propto \td t^2-\td\Omega_{n-2}^2\,,\quad t\in[0,2\pi)\,,
\end{equation}
which is an $(n-1)$-dimensional torus with $n-2$ temporal directions and one spatial direction, denoted by $T^{n-2,1} =S^{n-2} \times S^1$. In particular, for $n=3$, the dual field theory is a conformal field theory on a Lorentzian torus $T^{1,1}$.\footnote{The celestial CFT$_2$ dual to the bulk QFT in Klein space lives on the Lorentzian torus $T^{1,1}$. The correlation functions of the celestial CFT$_2$ are studied in \cite{Melton:2023hiq, Melton:2025ecj}.} However, in more general cases where $n\geq 4$, the dual field theory should be regarded as a multi-time QFT equipped with the symmetry $\mathrm{SO}(n-1,2)$, which could be regarded as the conformal symmetry of $T^{n-2,1}$. In the same spirit as ``flipped AdS$/\mathbb Z$'', we refer to this type of field theory as ``flipped CFT$_{n-1}$'' (or simply fCFT) for any $n \geq 3$. 

In this section, we develop the extrapolate dictionary for a free scalar field in asymptotically fAdS spacetimes in arbitrary dimensions. As an example, we will also calculate the holographic two-point functions at the globally flipped AdS/$\mathbb Z$ background. For a similar discussion of the holographic two-point function of globally flipped AdS$/\mathbb Z$ spacetime in three dimensions, refer to the recent work \cite{Fujiki:2026kca}. The relation to dS will be investigated in Sec. \ref{sec:kerrdS} as an application of the fAdS/fCFT correspondence.

\subsection{Generic discussions for extrapolate dictionary}

Let us first discuss the extrapolate dictionary, which requires us to study the asymptotic expansion of the fields $\phi$ near the spacetime boundary. Suppose the asymptotic behavior of $\phi$ is 
\begin{equation}
    \phi \underset{\rho \to\infty}{\longrightarrow} e^{- \rho \delta_\xi} \phi_\xi
\end{equation}
Then the equation of motion $(\nabla^2 -m^2) \phi=0$ at $\rho \rightarrow \infty$ gives
\begin{equation}
    [-\delta_\xi^2 + \delta_\xi (n-1) - m^2 l^2] e^{\rho (n-1)} \cdot (e^{-\delta_\xi \rho} \phi_\xi) = 0,
\end{equation}
where we expanded the equation in powers of $e^\rho$ and retained only the leading term. There are two solutions of $\delta_\xi$, which are $\delta_\pm$, given in \eqref{eq:SnCorrelator}. Then, the asymptotic expansion of $\phi$ is generally given by
\begin{equation} \label{eq:boundaryAdS}
   \phi (\rho, \vec y, t) \underset{\rho\to\infty}{\longrightarrow} \sum_{\xi = \pm} e^{- \rho \delta_\xi} \phi_\xi (\vec y, t),
\end{equation}
Note that the two modes $\phi_\pm$ are both normalizable at the boundary $\rho \to \infty$. The extrapolate dictionary is then proposed as 
\begin{equation}\label{eq:boundaryCFTdic}
    \hat \phi \underset{\rho\to\infty}{\longrightarrow} \sum_{\xi = \pm} e^{- \rho \delta_\xi} O_{\xi} (\vec y, t),
\end{equation}
where $O_\pm$ are operators with scaling dimensions $\delta_\pm$ of the flipped CFT$_{n-1}$ living on the boundary $\rho \to \infty$, and we add a hat to distinguish this operator-level dictionary from the asymptotic expansion \eqref{eq:boundaryAdS} for the classical solution. The form of the holographic two-point functions is fixed by conformal symmetry up to overall normalization. Consequently, the only non-vanishing two-point functions are $\<O_+ O_+\>$ and $\<O_- O_-\>$, whereas the mixed correlator $\< O_+ O_- \>$ vanishes because $O_+$ and $O_-$ have different scaling dimensions. For more details, refer to Appendix \ref{sec:flipCFT}.

It is important to remark that the holographic dictionary developed in this subsection applies to a general class of asymptotically fAdS spacetimes. Similar to the standard AdS/CFT correspondence, different bulk saddle points correspond to different states in the dual CFT. For example, the globally flipped AdS/$\mathbb Z$ background corresponds to the CFT vacuum, in which case the bulk Green's function $G(x';x)$ is given by \eqref{eq:FAdS2pt}, and the CFT correlators $\< O_\pm O_\pm\>$ are vacuum correlation functions. For other bulk backgrounds, such as the black hole solutions, the holographic correlators $\< O_\pm O_\pm\>$ are no longer vacuum correlators.

\subsection{Holographic dictionary in global fAdS}
As an explicit example, we apply the extrapolate dictionary to global fAdS$_n$ spacetime and compute the correlation function of flipped CFT$_{n-1}$ by taking the asymptotic limit of the bulk correlator in Eq.~\eqref{eq:FAdS2pt} near the conformal boundary. In this case, the asymptotic behavior \eqref{eq:boundaryAdS} can be checked by analyzing the asymptotic behavior of the modes in \eqref{eq:modeFAdS} after Wick rotation $\psi \rightarrow i \rho$. 

In the globally flipped AdS$/\mathbb Z$ background, the operator-valued boundary coefficients in Eq.~\eqref{eq:boundaryCFTdic} can be obtained directly from the asymptotic expansions of the $F$ and $G$ radial modes in Eq.~\eqref{eq:modeFAdS}, which are
\begin{equation}
O_{\xi}(\vec{\Omega},t)=\sum_{s,\{\nu\}}\sum_{k\in\mathbb Z} Y^{(n-1)}_{s,\{\nu\}}(\vec{\Omega})e^{ikt} \left[
f_{\xi;s,k}\,a^{(F)}_{s,\{\nu\};k}+g_{\xi;s,k}\,a^{(G)}_{s,\{\nu\};k} \right], \qquad \xi\in\{-,+\}.
\label{eq:boundary-mode-expansion-FG}
\end{equation}
For the continuation $\psi=\epsilon+i\rho$ with $\epsilon\to0^{+}$, the $G$ mode approaches the large-$z$ cut with $\arg(-z)=+\pi$, and the four asymptotic coefficients are
\begin{equation}
    \begin{aligned}
        f_{-;s,k}&=i^{s}2^{\delta_-}\frac{\Gamma(C)\Gamma(M)}{\Gamma(C-A)\Gamma(C-B)}, &
        f_{+;s,k}&=i^{s}2^{\delta_+}\frac{\Gamma(C)\Gamma(-M)}{\Gamma(A)\Gamma(B)}, \\
        g_{-;s,k}&=-\,i^{s}2^{\delta_-}e^{-i\pi a}\frac{\Gamma(a)\Gamma(M)}{2\Gamma(1+L)\Gamma(c-a)}, &
        g_{+;s,k}&=-\,i^{s}2^{\delta_+}e^{-i\pi b}\frac{\Gamma(b)\Gamma(-M)}{2\Gamma(1+L)\Gamma(c-b)},
\end{aligned}
\label{eq:boundary-mode-coefficients-FG}
\end{equation}
with
\begin{equation}
    \begin{aligned}
        A&=\frac{1+L-q-M}{2},&B&=\frac{1+L+q-M}{2},&C&=1+L, \\
        a&=\frac{1+q+L-M}{2},&b&=\frac{1+q+L+M}{2},&c&=1+q,
    \end{aligned}
\end{equation}
and $q\equiv |k|$ and $L\equiv s+(n-3)/2$.

The two boundary operators $O_-$ and $O_+$ are independent before a
quantum state is specified.  In the globally flipped
AdS$_n/\mathbb Z$ background, define $K_{\rm in}$ and $K_{\rm out}$
mode by mode as
\begin{equation}
\begin{aligned}
K_{\rm in}
\left[
Y^{(n-1)}_{s,\{\nu\}}(\vec\Omega)e^{ikt}
\right]
&=
\frac{f_{+;s,k}}{f_{-;s,k}}
Y^{(n-1)}_{s,\{\nu\}}(\vec\Omega)e^{ikt},
\\
K_{\rm out}
\left[
Y^{(n-1)}_{s,\{\nu\}}(\vec\Omega)e^{ikt}
\right]
&=
\frac{g_{+;s,k}}{g_{-;s,k}}
Y^{(n-1)}_{s,\{\nu\}}(\vec\Omega)e^{ikt}.
\end{aligned}
\label{eq:KinKout}
\end{equation}
Together with the Euclidean-cap conditions
\begin{equation}
a^{(G)}_{s,\{\nu\};k}\ket{0}=0,
\qquad
\bra{\pi/2}a^{(F)}_{s,\{\nu\};k}=0,
\end{equation}
the expansion \eqref{eq:boundary-mode-expansion-FG} gives
\begin{equation}
\boxed{
\left(O_+-K_{\rm in}O_-\right)\ket{0}=0,
\qquad
\bra{\pi/2}\left(O_+-K_{\rm out}O_-\right)=0.
}
\label{eq:shadow-state-relations}
\end{equation}
These are state conditions rather than operator identities on the full
Hilbert space.
For generic non-resonant $M$, the eigenvalues in
\eqref{eq:KinKout} satisfy the scalar shadow intertwining relations
\begin{equation}
K_\star\,\pi_{\delta_-}(g)
=
\pi_{\delta_+}(g)\,K_\star,
\qquad
\delta_+=n-1-\delta_-,
\qquad
\star\in\{{\rm in},{\rm out}\},
\label{eq:shadow-intertwiners}
\end{equation}
where $\pi_\delta$ denotes the scalar representation of
$\mathrm{SO}(n-1,2)$ with scaling dimension $\delta$.  Thus both
$K_{\rm in}$ and $K_{\rm out}$ are sector-wise scalar shadow
(Knapp-Stein) intertwiners.  They need not coincide, since the two
Euclidean caps select different analytic branches and sector
normalizations.

For holographic correlation functions, we first take $\rho\rightarrow \infty$ in \eqref{eq:FAdS2pt} to obtain
\begin{equation} \label{eq:asyAdS2pt}
    \begin{split}
        & G (\rho, \vec y, t; \rho', \vec y', t') = \< \phi(x) \phi(x')\>_{\text{flipped AdS}/\mathbb{Z}} \\
        & \underset{\rho \to \infty}{\longrightarrow} \frac{l^{2-n}}{4 \pi^{\frac{n+1}{2}}} \sum_{\xi = \pm} \frac{2^{\delta_\xi} \Gamma(\delta_\xi) \Gamma(\frac{n-1}{2} - \delta_\xi) e^{-(\rho+ \rho') \delta_\xi}}{[y \cdot y' - \cos (t-t') - (y \cdot y' + \cos (t-t')) e^{-2 \rho'}+i\epsilon]^{\delta_\xi}} \,,
    \end{split}
\end{equation}
from which we can read out two bulk-to-boundary propagators, 
\begin{equation}
    \<O_\pm(t,y^i)\phi(x')\>= \frac{2^{\delta_\pm-2} l^{2-n} \pi^{-\frac{n+1}{2}} \Gamma(\delta_\pm) \Gamma(\frac{n-1}{2} - \delta_\pm) e^{- \rho' \delta_\pm}}{[y \cdot y' - \cos (t-t') - (y \cdot y' + \cos (t-t')) e^{-2 \rho'}+i\epsilon]^{\delta_\pm}}\,,
\end{equation}
Then we can take $\rho'\to\infty$ to obtain the boundary flipped CFT$_{n-1}$ correlation function. When $y \cdot y' \neq \cos (t-t')$, the limits $\rho, \rho' \to \infty$ can be taken simultaneously, yielding 
\begin{equation} \label{eq:GAdSInfty1}
    G (\rho, \vec y, t; \rho', \vec y', t') = 
    \underset{\rho,\rho' \to\infty}{\longrightarrow} \frac{l^{2-n}}{4 \pi^{\frac{n+1}{2}}} \sum_{\xi = \pm} e^{-(\rho+ \rho') \delta_\xi} \frac{2^{\delta_\xi} \Gamma(\delta_\xi) \Gamma(\frac{n-1}{2} - \delta_\xi)}{(y \cdot y' - \cos (t-t')+i\epsilon)^{\delta_\xi}} 
\end{equation}
However, when $y \cdot y' = \cos (t-t')$, \eqref{eq:asyAdS2pt} becomes
\begin{equation} \label{eq:GAdSInfty2}
    G (\rho, \vec y, t; \rho', \vec y', t') \underset{\rho \to \infty}{\longrightarrow} \frac{l^{2-n}}{4 \pi^{\frac{n+1}{2}}} \sum_{\xi =\pm} e^{-(\rho- \rho') \delta_\xi} \frac{\Gamma(\delta_\xi) \Gamma(\frac{n-1}{2} - \delta_\xi)}{(-\cos (t-t')+i\epsilon)^{\delta_\xi}}\,,
\end{equation}
where the flipped CFT$_{n-1}$ correlation functions can be read out only if $\rho-\rho'\to\infty$.

In conclusion, the flipped CFT$_{n-1}$ correlation functions have two branches, one of which is
\begin{equation}\label{eq:powerlawBranch}
    \<O_\pm (t,y^i) O_\pm (t',y'^i)\>= \frac{c_\pm}{(y \cdot y' - \cos (t-t')+i\epsilon)^{\delta_\pm}} \,.
\end{equation}
This is the standard form of a two-point function of flipped CFT on $T^{n-2,1}$ with metric \eqref{eq:T1n-2}. This branch of the holographic two-point function can be recovered in the context of the flipped CFT$_{n-1}$ by conformal symmetries, see the detailed discussions in Appendix \ref{sec:flipCFT}. 
Moreover, the extrapolate dictionary implies that the mixed two-point function $\langle O_- O_+\rangle$ vanishes for two generically separated points. This is also consistent with the symmetry analysis of the flipped CFT, since the conformal symmetry requires the two-point function between operators with different conformal dimensions to vanish. 

The other branch is
\begin{equation}
    \<O_\pm (t,y^i) O_\pm (t',y'^i)\> \propto \frac{1}{(\cos (t-t')-i\epsilon)^{\delta_\pm}} \,, ~ \text{with}~ y\cdot y'-\cos(t-t')=0.
\end{equation}
This branch resembles the delta function branch \cite{Bagchi:2022emh, Mason:2023mti, Alday:2024yyj, Nguyen:2025sqk} of the two-point function in Carrollian CFT, which is central in flat holography. However, this branch only has support on the light-cone, which is not the case of interest in the present work. Therefore, we do not include this branch in the following analysis.

\section{Entropy of dS$_3$ and Kerr-dS$_3$ from flipped CFT$_2$} \label{sec:kerrdS}

For a two-dimensional CFT on a Euclidean torus, the asymptotic density of states at large conformal weights is
given by the Cardy formula
\begin{equation}
 \log \rho(h,\bar h) \simeq 2\pi \sqrt{\frac{c_{\mathrm{L}}}{6} \left(h-\frac{c_{\mathrm{L}}}{24}\right)} + 2\pi  \sqrt{\frac{c_{\mathrm{R}}}{6} \left(\bar h-\frac{c_{\mathrm{R}}}{24} \right)},
\end{equation}
where $c_{\mathrm{L}}$ and $c_{\mathrm{R}}$ are the left- and right-moving central charges.
Equivalently, in the canonical ensemble with left- and right-moving inverse
temperatures $\beta_{\mathrm{L}}$ and $\beta_{\mathrm{R}}$, the high-temperature partition function with the modular parameter $(\tau, \bar \tau) = \frac{i}{2\pi} (\beta_{\mathrm{L}}, -\beta_{\mathrm{R}})$
behaves as
\begin{equation}
Z(\beta_{\mathrm{L}},\beta_{\mathrm{R}}) = \mathrm{Tr}\left[ e^{2\pi i\tau(L_0-\frac{c}{24})} e^{-2\pi i\bar\tau(\bar L_0-\frac{\bar c}{24})} \right] \simeq e^{\frac{\pi^2 c_{\mathrm{L}}}{6\beta_{\mathrm{L}}} + \frac{\pi^2 c_{\mathrm{R}}}{6\beta_{\mathrm{R}}}},
\end{equation}
and the corresponding thermal entropy is
\begin{equation}  \label{eq:Cardydef}
S=\frac{\pi^2}{3}\left(c_{\mathrm{L}} T_{\mathrm{L}}+c_{\mathrm{R}} T_{\mathrm{R}} \right), \qquad T_{\mathrm{L,R}}=\beta_{\mathrm{L,R}}^{-1}.
\end{equation}
The fCFT$_2$ considered here is instead defined on a Lorentzian torus. Similar to the recent discussion in Ref.~\cite{Fujiki:2026kca}, we use \eqref{eq:Cardydef} with the central charges and modular parameters analytically continued to those of the Lorentzian torus. We will show that the resulting Cardy entropy reproduces the cosmological-horizon entropy for both pure dS$_3$ and Kerr-dS$_3$.

\subsection{Entropy of dS$_3$} \label{app:dS3Entropy}

Let us begin with the entropy of the cosmological horizon of dS$_3$. Recall that the static-patch metric of dS$_3$ in our conventions is
\begin{equation}
    \td s^2/l^2 = -(1 - r^2 ) \td t_*^2 + \frac{\td r^2}{1- r^2} + r^2 \td \phi^2\,.
\end{equation}
The cosmological horizon is defined at $r= 1$. The near-horizon geometry can be rewritten as a Rindler space by $R = \sqrt{2(1-r)}$
\begin{equation}
    \td s^2/l^2 \approx - R^2 \td t_*^2 + \td R^2 + \td \phi^2\,,\quad R\ll 1\,.
\end{equation}
To avoid the conical singularity, $t = i t_*$ needs to have a period $2\pi$, which can be interpreted as the inverse of the Gibbons--Hawking temperature, i.e., $T = \frac{1}{2 \pi}$. The thermodynamic entropy follows from the area $A_{\text{cosmology}}=2\pi l$ of the cosmological horizon, 
\begin{equation}
    S=\frac{A_{\text{cosmology}}}{4G}=\frac{\pi l}{2 G}\,.
\end{equation}
In the literature, this entropy can alternatively be derived through the dS$_3$/CFT$_2$ correspondence by naively using the Cardy formula
\begin{equation}
    S= \frac{2\pi^2}{3}c \, T = \frac{\pi l}{2G},
\end{equation}
where the dual CFT$_2$ with the central charge $c = \frac{3l}{2G}$ is defined on the past spacelike conformal boundary $\mathcal{I}^-$ of dS$_3$, as proposed by \cite{Strominger:2001pn}.

In the flipped AdS$_3/\mathbb{Z}$ spacetime with metric
\begin{equation}
    \td s^2/l^2=-\td \rho^2-\sinh^2\rho \td\phi^2+\cosh^2\rho\td t^2\,,\quad\rho\in[0,\infty), ~ t, \phi\in[0,2\pi) \,,
\end{equation}
we can consider its dual CFT$_2$ living on the conformal boundary with metric
\begin{equation}
    \td s^2\propto \td t^2-\td\phi^2\,,\quad\phi\in[0,2\pi),\quad t\in[0,2\pi)\,,
\end{equation}
which is a Lorentzian torus. The partition function of a CFT$_2$ on a Lorentzian torus can be obtained by analytic continuation of the Cardy formula. We regard the inverse period of the spacelike coordinate $t$ as an imaginary temperature $T=\frac{1}{2\pi i}$. And with an imaginary central charge $c=\frac{3l_{\text{AdS}}}{2G}=\frac{3il}{2G}$, the Cardy formula \eqref{eq:Cardydef} gives 
\begin{equation} \label{eq:dSEntropy}
    S=\frac{2\pi^2}{3}cT = \frac{\pi l}{2G}\,,
\end{equation}
which reproduces the entropy of dS$_3$.

\subsection{Entropy of Kerr-dS$_3$}

Kerr-dS$_3$  is obtained from global dS$_3$ as a quotient by a discrete group action. This operation admits an analytic continuation to $S^3$, which is realized as a quotient of $S^3$ by the corresponding discrete group, giving rise to a conical defect geometry. The subsequent analytic continuation yields the associated asymptotically flipped AdS/$\mathbb{Z}$ spacetime, together with the corresponding modular parameter of the dual flipped CFT.

We first consider the case with vanishing angular momentum $J=0$, where $r_-=0$ but $r_+<l$. The associated Euclidean spacetime should be a three-dimensional sphere with a single one-dimensional conical defect, which we refer to as a conifold $S^3$.

The metric of the conifold $S^3$ can be written in two different ways. One is
\begin{equation}
    \td s^2/l^2=d\theta_1^2+\cos^2\theta_1 \left(\td\theta_2^2+\left(\frac{r_+}{l}\right)^2\cos^2\theta_2 \td\phi^2\right)\,,\quad \theta_1,\theta_2\in\left[-\frac{\pi}{2},\frac{\pi}{2}\right]\,,\quad\phi\in[0,2\pi)\,,
\end{equation}
with a single closed conical defect, that is, the circle passing through the points ($A,B,D$) in Fig.~\ref{fig:hemi-S2}. This conical defect can be divided into two parts, which are denoted by $\theta_2=\pm\frac{\pi}{2}$. These two parts are connected at the points $\theta_1=\pm\frac{\pi}{2}$.
The other is
\begin{equation}\label{eq:ConicalS3}
    \td s^2/l^2=d\psi^2+\left(\frac{r_+}{l}\right)^2\sin^2\psi \td\phi^2+\cos^2\psi\td t^2\,,\quad \psi\in\left[0,\frac{\pi}{2}\right]\,,\quad t,\phi\in[0,2\pi)\,,
\end{equation}
with the single closed conical defect denoted by $\psi=0$.

Analytic continuation $\theta_1\rightarrow i\tau$ gives us the Kerr-dS spacetime with $J$=0 
\begin{equation}\label{eq:Sch-dS}
    \td s^2/l^2=-d\tau^2+\cosh^2\tau\left(\td\theta_2^2+\left(\frac{r_+}{l}\right)^2\cos^2\theta_2 \td\phi^2\right)\,,\quad \tau\in(-\infty,\infty)\,,
\end{equation}
with two disconnected conical defects. These two conical defects can be regarded as two point-like spinless massive objects travelling through dS spacetime at antipodal positions $\theta_2=\pm\frac{\pi}{2}$.

Analytic continuation $\psi\rightarrow i\rho$ gives us a flipped AdS$_3/\mathbb{Z}$ manifold with a single spacelike conical defect at $\rho=0$, whose metric takes the following form: 
\begin{equation}
    \td s^2/l^2=-\td\rho^2-\left(\frac{r_+}{l}\right)^2\sinh^2\rho \td\phi^2+\cosh^2\rho\td t^2\,.
\end{equation}
We may consider a dual CFT$_2$ living on the conformal boundary $\rho\rightarrow\infty$,
\begin{equation}
    \td s^2\propto \left(\frac{l}{r_+}\right)^2\td t^2-\td\phi^2\,,
\end{equation}
which is a Lorentzian torus corresponding to an imaginary temperature $T=\left(\frac{i2\pi l}{r_+}\right)^{-1}$. Again, employing the Cardy formula gives the entropy
\begin{equation}
    S=\frac{2\pi^2}{3} c T=\frac{2\pi r_+}{4G}\,.
\end{equation}

For the Kerr-dS$_3$ case with mass $\mathcal{M}$ and nonvanishing angular momentum $J$, the associated Euclidean spacetime would have a complex metric near the conical defect. Away from the conical defect, it still behaves like a conifold $S^3$, whose metric is
\begin{equation}
    \td s^2=N^{-2}\td r^2+r^2(\td\phi+i N^\phi\td t)^2+N^2\td t^2\,,\quad r\in[0,r_+]
\end{equation}
with $r=r_+ \sin\psi$ and
\begin{equation}
    N^2 = \frac{(r_+^2-r^2) (r^2 + r_-^2)}{l^2 r^2} = \mathcal{M} - \frac{r^2}{l^2} + \frac{J^2}{4r^2}, \quad N^\phi = \frac{r_+ r_-}{l r^2} = \frac{J}{2 r^2}.
\end{equation}
Or equivalently,
\begin{equation}
    N^2=\frac{r_+^2}{l^2} \cos^2\psi \left(1+\frac{r_-^2}{r_+^2}\csc^2\psi\right)\,,\quad N^\phi=\frac{\Omega_{\textrm{\tiny H}}}{l\sin^2\psi}
\end{equation}
and
\begin{equation}
    \td s^2/l^2=\frac{\td \psi^2}{1+\OmegaH^2\csc^2\psi}+\frac{r_+^2}{l^2}\sin^2\psi(\td\phi+i N^\phi\td t)^2+\frac{r_+^2}{l^4} \cos^2\psi (1+\OmegaH^2\csc^2\psi)\td t^2
\end{equation}
with the angular velocity of the cosmological horizon $\OmegaH=\frac{r_-}{r_+}$. 
Finally, by rescaling the coordinate 
\begin{equation}\label{eq:rescaling}
    t\rightarrow \frac{l^2/r_+}{1+\OmegaH^2}t\,,
\end{equation}
the metric becomes
\begin{equation} \label{eq:S3dSconifold}
    \begin{split}
        \td s^2/l^2= &\frac{\td \psi^2}{1+\OmegaH^2\csc^2\psi} +\cos^2\psi \frac{1+\OmegaH^2\csc^2\psi}{(1+\OmegaH^2)^2}\td t^2\\
        & + \frac{r_+^2}{l^2}\sin^2\psi\left(\td\phi+i \frac{\OmegaH}{1+\OmegaH^2} \frac{l/r_+}{\sin^2\psi}\td t\right)^2
    \end{split}
\end{equation}
which returns to \eqref{eq:ConicalS3} if $\OmegaH=0$. The near horizon geometry $r \approx r_+$ can be rewritten by $\psi=\frac{\pi}{2} - \sqrt{1+\OmegaH^2} \epsilon ~(\epsilon \ll 1)$ as
\begin{equation}
    \td s^2/l^2 = \td \epsilon^2 + \epsilon^2 \td t^2 + \frac{r_+^2}{l^2} \left(\td\phi+i \frac{\OmegaH}{1+\OmegaH^2} l/r_+ \td t\right)^2.
\end{equation}
Then, the regularity at $\psi=\frac{\pi}{2}$ requires that
\begin{equation}\label{eq:periodKerrdS}
    t\simeq t+2\pi\,,\quad \phi\simeq\phi-i \frac{2\pi\OmegaH}{1+\OmegaH^2} l/r_+\,.
\end{equation}
We can also use the reparameterization \eqref{eq:reparameterization} of a hemi-$S^2$ to rewrite this manifold as
\begin{equation}
    \begin{aligned}
        \td s^2/l^2&=\frac{(\cos\theta_2\sin\theta_1\td\theta_1+\cos\theta_1\sin\theta_2\td\theta_2)^2}{(1-\cos^2\theta_1\cos^2\theta_2)(1+\OmegaH^2 \sec^2\theta_1\sec^2\theta_2)}\\
        & +\frac{(\sin\theta_1\cos\theta_1\cos\theta_2\td\theta_2-\sin\theta_2\td\theta_1)^2(1+\OmegaH^2 \sec^2\theta_1\sec^2\theta_2)}{(1-\cos^2\theta_1\cos^2\theta_2)(1+\OmegaH^2)^2}\\
        &+\frac{\left(i\OmegaH(\sec\theta_1\tan\theta_2\td\theta_1-\sin\theta_1 \td\theta_2)+\frac{r_+}{l}(1+\OmegaH^2)\cos\theta_1\cos\theta_2(1-\cos^2\theta_1\cos^2\theta_2)\td\phi\right)^2}{(1+\OmegaH^2)^2(1-\cos^2\theta_1\cos^2\theta_2)^2}
        \,.
    \end{aligned}
\end{equation}

Analytic continuation $\theta_1\rightarrow i\tau$ gives an asymptotically dS$_3$ spacetime, which is in fact a Kerr-dS$_3$ spacetime. Its metric is given by
\begin{equation}
    \begin{aligned}
        \td s^2/l^2&=\frac{(-\cos\theta_2\sinh\tau\td\tau+\cosh\tau\sin\theta_2\td\theta_2)^2}{(1-\cosh^2\tau\cos^2\theta_2)(1+\OmegaH^2 \cosh^{-2}\tau\sec^2\theta_2)}\\
        & -\frac{(\sinh\tau\cosh\tau\cos\theta_2\td\theta_2-\sin\theta_2\td\tau)^2(1+\OmegaH^2 \cosh^{-2}\tau\sec^2\theta_2)}{(1-\cosh^2\tau\cos^2\theta_2)(1+\OmegaH^2)^2}\\
        &+\frac{\left(\OmegaH (-\text{sech}\,\tau \tan\theta_2\td\tau+\sinh\tau \td\theta_2)+\frac{r_+}{l}(1+\OmegaH^2)\cos\theta_2(1-\cosh^2\tau\cos^2\theta_2)\cosh\tau\td\phi\right)^2}{(1+\OmegaH^2)^2(1-\cosh^2\tau\cos^2\theta_2)^2}
        \,,
    \end{aligned}
\end{equation}
which returns to \eqref{eq:Sch-dS} if $\OmegaH=0$.

Analytic continuation $\psi\rightarrow i\rho$ gives an asymptotically flipped AdS/$\mathbb{Z}$ spacetime with metric,
\begin{equation}
    \begin{split}
        \td s^2/l^2=&-(1-\OmegaH^2/\sinh^2\rho)^{-1}\td \rho^2 +\cosh^2\rho \frac{1-\OmegaH^2/\sinh^2\rho}{(1+\OmegaH^2)^2}\td t^2\\
        &-\frac{r_+^2}{l^2}\sinh^2\rho\left(\td\phi-i \frac{\OmegaH}{1+\OmegaH^2} \frac{l/r_+}{\sinh^2\rho}\td t\right)^2
    \end{split}\label{asymflipped}
\end{equation}
Then the dual CFT$_2$ can be defined on the conformal boundary $\rho\rightarrow\infty$,
\begin{equation}
    \td s^2/l^2\propto \frac{l^2}{r_+^2(1+\OmegaH^2)^2}\td t^2-\td\phi^2= -\td w \td \bar w \,,
\end{equation}
where $(w, \bar w) = (\phi - \frac{l/r_+}{1+\OmegaH^2} t, \phi + \frac{l/r_+}{1+\OmegaH^2} t)$. Using the periodic identification $\phi\simeq\phi+2\pi$ together with Eq.~\eqref{eq:periodKerrdS}, the Lorentzian torus has the modular parameter
\begin{equation} \label{eq:modular}
    \tau =-\frac{l}{r_+ - ir_-}\,, \qquad \bar{\tau} = \frac{l}{r_+ + ir_-}\,.
\end{equation}
Another way to derive the metric \eqref{asymflipped} and the modular parameters $\tau, \bar{\tau}$ is to perform the coordinate transformation on $S^3$ and then implement analytic continuation. For more details, refer to Appendix \ref{sec:conifoldS3}. Finally, the Cardy formula gives
\begin{equation}
    S=\frac{i\pi c}{6}(1/\tau -1/\bar{\tau}) =\frac{\pi^2 c}{3}(T_{\mathrm{L}}+T_{\mathrm{R}})=\frac{2\pi r_+}{4G}
\end{equation}
with
\begin{equation}
    T_{\mathrm{L}}=\frac{i}{2\pi \tau}=\frac{r_+-ir_-}{2\pi i l}\,, \qquad T_{\mathrm{R}}=\frac{r_++ir_-}{2\pi i l}\,.
\end{equation}

\section{Conclusion and discussions}
A central result of this work is the analytic relation between quantum field theories in dS and flipped $\mathrm{AdS}/\mathbb{Z}$. This relation originates from their common Euclidean preparation. The correlation functions considered here are defined with respect to the Euclidean vacuum, while both Lorentzian spacetimes arise from distinct analytic continuations of the same Euclidean sphere $S^n$. In particular, dS$_n$ is obtained through the continuation $\theta_1\rightarrow i\tau$, whereas the alternative continuation $\psi\rightarrow i\rho$ leads to flipped $\mathrm{AdS}_n/\mathbb{Z}$, a multi-time spacetime whose constant-$\rho$ slices are compact.

At the level of the Euclidean path integral, the vacuum partition functions of QFTs in pure dS and pure flipped $\mathrm{AdS}/\mathbb{Z}$ are therefore obtained from the same path integral on $S^n$. More generally, correlation functions in the two Lorentzian theories originate from the same Euclidean correlation functions and differ only in the analytic continuation used to reach the corresponding Lorentzian geometry and in the resulting operator-ordering prescription. In dS, the Lorentzian correlators are ordered with respect to the physical time $\tau$, whereas in flipped $\mathrm{AdS}/\mathbb{Z}$ they are ordered with respect to the radial evolution parameter $\rho$. This common Euclidean origin provides a direct analytic continuation between correlation functions in the two theories.

We further confirm this relation through an explicit canonical quantization. Two different Hamiltonian foliations of the Euclidean theory on $S^n$ can be analytically continued, under the corresponding Wick rotations, to the canonical formulations of QFT in dS$_n$ and flipped $\mathrm{AdS}_n/\mathbb{Z}$, respectively. Although the two Euclidean foliations define different evolution schemes, they lead to the same Euclidean propagator. Their respective analytic continuations therefore yield the Feynman propagators in dS$_n$ and fAdS$_n$, establishing explicitly that the two Lorentzian correlators can be mapped into one another through the corresponding analytic continuations.

The relation between QFT in dS spacetime and QFT in fAdS closely parallels the previously established relation between QFT in Minkowski spacetime and QFT in Klein space \cite{Chen:2025acl,Chen:2025eeh}. In both cases, the two Lorentzian theories arise from distinct analytic continuations of a common maximally symmetric Euclidean geometry: the sphere $S^n$, with isometry group $\mathrm{SO}(n+1)$, in the dS/fAdS case, and Euclidean space, with isometry group $\mathrm{ISO}(n)$, in the Minkowski/Klein case. This parallel is also reflected in their asymptotic structures. Minkowski spacetime with isometry group $\mathrm{ISO}(1,n-1)$ possesses distinct future and past null infinities, while dS spacetime with $\mathrm{SO}(1,n)$ has two disconnected conformal boundaries $\mathcal{I}^{\pm}$. By contrast, Klein space with $\mathrm{ISO}(2,n-2)$ and fAdS spacetime with $\mathrm{SO}(n-1,2)$ each possess a single connected conformal boundary.

A further important observation is that, under the analytic continuation
$l\rightarrow -i l_{\mathrm{AdS}}$, fAdS spacetime is locally continued to ordinary AdS spacetime. This close relation to AdS naturally motivates us to seek a holographic description of QFT in fAdS spacetime in analogy with the standard AdS/CFT correspondence. Combined with the analytic relation between fAdS and dS established above, such a holographic description may in turn provide an alternative route to toward de Sitter holography. We therefore propose that QFT in flipped $\mathrm{AdS}_n/\mathbb{Z}$ admits a dual description in terms of a boundary theory that we refer to as flipped CFT$_{n-1}$ (fCFT$_{n-1}$), defined on the Lorentzian torus forming the conformal boundary of fAdS. In this work, we determine its two-point functions in two complementary ways. From the bulk perspective, we extract them from the asymptotic behavior of bulk two-point functions using the extrapolate dictionary. Independently, from the boundary perspective, we derive the corresponding vacuum two-point functions directly from the conformal symmetry of fCFT. The agreement between these two constructions provides a nontrivial consistency check of the proposed fAdS/fCFT correspondence.

Using the holographic description of flipped $\mathrm{AdS}/\mathbb{Z}$ developed in this work, we can revisit some of the subtleties encountered in the conventional dS/CFT correspondence. Through the analytic continuation relating $\mathrm{dS}_n$ and fAdS$_n$, the fCFT$_{n-1}$ living on the Lorentzian-torus boundary of fAdS$_n$ provides a natural candidate for encoding de Sitter observables. In particular, the Lorentzian torus admits a natural interpretation of the imaginary temperature that appears in the dS$_3$/CFT$_2$ analysis, while the corresponding imaginary central charge is already familiar from the conventional dS/CFT literature. In three dimensions ($n=3$), these ingredients combine in a particularly simple way. Using the Cardy formula of fCFT$_2$, we reproduce the entropy of the dS$_3$ cosmological horizon, in exact agreement with the Bekenstein--Hawking entropy obtained from its horizon area. This agreement provides a nontrivial indication that the fCFT$_2$ captures microscopic information associated with the de Sitter cosmological horizon. The construction extends naturally to asymptotically dS$_3$ geometries, in particular to Kerr-dS$_3$. The corresponding Euclidean geometry can be described as an $S^3$ geometry with a conical defect, from which both Kerr-dS$_3$ and its asymptotically fAdS$_3/\mathbb{Z}$ counterpart can be obtained by appropriate analytic continuations. Equivalently, as shown in Appendix~\ref{sec:conifoldS3}, these geometries may be constructed as quotients by a discrete group action: starting from pure dS$_3$ and pure fAdS$_3$, one implements the same discrete group action on the two sides to generate the corresponding quotient spacetimes. This construction allows the modular parameter of the associated fCFT$_2$ to be identified and, once again, the Cardy formula reproduces the Bekenstein--Hawking entropy of the Kerr-dS$_3$ cosmological horizon.

In the present work, we restrict this analysis to three bulk dimensions. Extending the construction to higher-dimensional de Sitter spacetimes is an important direction for future investigation. A new feature arises for Kerr-dS$_n$ with $n\geq4$, where a black-hole event horizon appears in addition to the cosmological horizon. Understanding how these multiple horizons and their associated entropies are encoded in the proposed fAdS/fCFT framework will require a more detailed analysis.

As we have shown, a formal analytic continuation of the Cardy formula reproduces the thermodynamic entropy of the de Sitter cosmological horizon. Nevertheless, its microscopic interpretation remains subtle. The boundary dual field theory describing dS bulk spacetime is generally non-unitary \cite{Balasubramanian:2002zh,Kawamoto:2023nki,Doi:2024nty},\footnote{Related issues also arise in flat-space holography; see, for example, Refs.~\cite{Bagchi:2019unf,Crawley:2021ivb}, and in particular the detailed discussion in Sec.~3.3.1 of Ref.~\cite{Kulp:2024scx}.} as already emphasized in early studies of dS/CFT \cite{Strominger:2001pn,Bousso:2001mw}. It therefore remains an important question whether, and under what conditions, the usual Cardy formula continues to apply in such a non-unitary setting.

A natural direction for future work is to determine whether the analytic continuation relating flipped $\mathrm{AdS}_{d+1}/\mathbb{Z}$ and $\mathrm{dS}_{d+1}$ induces a corresponding analytic continuation on the boundary side. Since fAdS$_{d+1}$ admits a holographic description in terms of the signature-flipped $\mathrm{CFT}_{d}$ developed in this work, establishing such an induced boundary map $\mathcal{A}_{\rm bdry}$ (as defined in Fig.~\ref{fig:analytic-continuation-dS-holography}) could provide  a new route toward constructing an alternative holographic description of de Sitter spacetime.

Recall that the conformal boundary of fAdS$_{d+1}$ is the Lorentzian torus $T^{d-1,1} \equiv S^{d-1}\times S^1$. \footnote{Recall that the boundary metric for fCFT is given by the signature-flipped metric, 
	\begin{equation}
		ds_{\rm bdry}^{2} \propto dt^2-d\Omega_{d-1}^2,
		\qquad
		t\sim t+2\pi \,.
\end{equation}}
The main open question is then to determine how the bulk analytic continuation acts on this boundary theory,
\begin{equation}
	\mathrm{fCFT}_{d}[T^{d-1,1}]
	\xrightarrow{\ \mathcal{A}_{\rm bdry}\ }
	\mathrm{QFT}_{d}^{\mathrm{dS}}[\mathcal{B}_{d}^{\mathrm{dS}}],
	\label{eq:induced-boundary-map}
\end{equation}
where both the target theory $\mathrm{QFT}_{d}^{\mathrm{dS}}$ and the $d$-dimensional background $\mathcal{B}_{d}^{\mathrm{dS}}$ on which it is defined are, at this stage, to be determined. There are at least two natural possibilities.

\paragraph{Route A: recovery of conventional dS/CFT.}

The first possibility is that the induced continuation ultimately reproduces
the conventional dS/CFT correspondence. For global $\mathrm{dS}_{d+1}$, the future and past conformal boundaries have topology $\mathcal{I}^{\pm}\simeq S^d $ and the conventional dS/CFT proposal associates de Sitter gravity with a Euclidean $\mathrm{CFT}_d$ defined on $S^d$. In this case,  the required boundary continuation would schematically take the form
\begin{equation}
	\boxed{
		\mathrm{fCFT}_{d} 	\left[	S^{d-1}\times S^1	\right]
		\underset{\scriptstyle ?}{\xrightarrow{\ \mathcal{A}_{\rm bdry}\ }}
		\mathrm{CFT}_{d}^{\mathrm{dS}} \left[S^d	\right]\,. 
	}
	\label{eq:route-A}
\end{equation}
An important difference from the standard EAdS/dS continuation is that the two boundary theories are not naturally defined on the same real boundary geometry. The relation between the two real boundary geometries of flipped AdS and dS is intrinsically more involved. The continuation may also involve an appropriate complexification and choice of real slice or contour \footnote{A simple indication of this nontriviality follows from the coordinate relation between the two parameterizations of the Euclidean sphere, $\sin\psi=\cos\theta_{1}\cos\theta_{2}$. The dS and flipped-AdS geometries are obtained from the different analytic continuations $\theta_{1}\rightarrow i\tau$ and $\psi\rightarrow i\rho$, respectively. Formally combining the two gives
$i\sinh\rho=\cosh\tau\cos\theta_{2}$. Hence, for generic real parameters $\rho$, $\tau$, and $\theta_{2}$, the two sides have incompatible reality properties. This indicates that a generic real point of flipped AdS does not map directly to a generic real point of dS, and suggests that the corresponding boundary continuation should involve a complexification together with an appropriate choice of real slice or contour.}. If this construction can be established at the level of correlation functions and generating functionals, the flipped-AdS description would provide an alternative derivation of conventional dS/CFT.

\paragraph{Route B: a different realization of dS holography.}

A second possibility is that the analytic continuation induced by the bulk relation leads to a $d$-dimensional QFT defined on an analytically continued version of the Lorentzian torus,
\begin{equation}
	\boxed{ \mathrm{fCFT}_{d}[T^{d-1,1}]
		\underset{\scriptstyle ?}{\xrightarrow{\ \mathcal{A}_{\rm bdry}\ }}
	\mathrm{QFT}_{d}^{\mathrm{dS}} [\widetilde{T}^{\,d-1,1}] \,.}
	\label{eq:route-B}
\end{equation}
Here $\widetilde{T}^{\,d-1,1}$ denotes the boundary geometry selected by the analytic continuation and need not coincide exactly with the original $T^{d-1,1}$. It would represent a realization of dS holography that is
different from, or possibly complementary to, the conventional Euclidean dS/CFT formulation.

It would therefore be particularly interesting to determine $\mathcal{A}_{\rm bdry}$ explicitly and to understand which of these two possibilities is realized. The two-point functions obtained in this work provide a natural starting point for this investigation. One may study how their operator dimensions, analytic structure, and $i\epsilon$-prescriptions transform under the continuation, and then ask whether the resulting boundary correlators coincide with those of conventional dS/CFT or instead define a different holographic boundary theory.

A broader possibility is that the two routes discussed above may admit a unified interpretation in terms of a parent complexified boundary theory. This would closely parallel the bulk construction developed in this work. Indeed, fAdS$_{d+1}$ and $\mathrm{dS}_{d+1}$ can be viewed as different real slices obtained by analytic continuation from a common complexified geometry with a conformal group $SO(d+2, \mathbb{C})$. It is therefore natural to ask whether an analogous structure exists on the boundary side.

More specifically, instead of regarding the flipped CFT and the putative dS dual as two distinct theories, we speculate that they arise as different real or contour realizations of a single complexified boundary theory, schematically
\begin{equation*}
	\boxed{
		\begin{array}{ccccc}
			&&
			 \mathrm{QFT}_{\mathbb{C}}
			&&
			\\[8pt]
			&
			\swarrow
			&&
			\searrow
			&
			\\[8pt]
			\mathrm{fCFT}_{d}[T^{d-1,1}]
			&&&&
			\mathrm{QFT}_{d}^{\mathrm{dS}}[\mathcal{B}_{d}^{\mathrm{dS}}]
		\end{array}
	}
\end{equation*}
In this picture, the induced continuation $\mathcal{A}_{\rm bdry}$ would correspond to moving between different real slices or contour prescriptions of the same underlying complexified field theory. The two possible endpoints discussed above could then correspond to different choices of this boundary realization. Establishing such a picture would require identifying the appropriate complexified boundary theory and showing explicitly how its correlation functions and generating functional reduce to the fCFT and dS quantities under different choices of real slice or contour.

\acknowledgments
We thank Yu-Ting Wen for valuable discussions and Feng Hao, Hao Ouyang, Xi-Yang Ran, Jie Xu, Zhijun Yin, and Zi-Xuan Zhao for valuable comments on the manuscript. This research is supported in part by NSFC Grants No.~12275004 and No.~12588101. SMR is supported by Peking University under startup Grant No. 7101303985.

\appendix

\section{QFT in dS$_n$ spacetime from $S^n$} \label{sec:dSQFT}

The dS spacetime with metric \eqref{eq:metricdS} has two conformal boundaries $\mathcal{I}^\pm: \tau\rightarrow\pm\infty$, which means the QFT defined on dS spacetime admits an S-matrix giving transitions among asymptotic states defined on $\mathcal{I}^\pm$ just as in Minkowski spacetime.

Several classes of vacuum states have been discussed in the literature. Generally, there exists a one-parameter family of vacua with dS invariance, which are called $\alpha$-vacua and labeled as $|\alpha\>$ \cite{Allen:1985ux, deBoer:2004nd}. In our framework, the correlation functions are expected to be connected with those in $S^n$ with the substitution $\theta_1 \rightarrow i \tau$ for the fields. Thus, we focus on the Euclidean vacuum $|0\>$, which is a special case of $\alpha$-vacuum $|\alpha=0\>$. The term ``Euclidean vacuum" refers to a vacuum state prepared by a Euclidean path integral. In de Sitter spacetime, the Euclidean vacuum is also known as the Bunch-Davies (BD) vacuum.

\begin{figure}[htbp]
    \centering
    \subfigure[The $\tau$-ordered $n$-point correlation function $\bra{0} \phi_n (x_n) \cdots \phi_1 (x_1) \ket{0}$, where $\tau_1 < \cdots < \tau_n$. \label{fig:pathCorretheta}]{
    \includegraphics[width=0.45\linewidth]{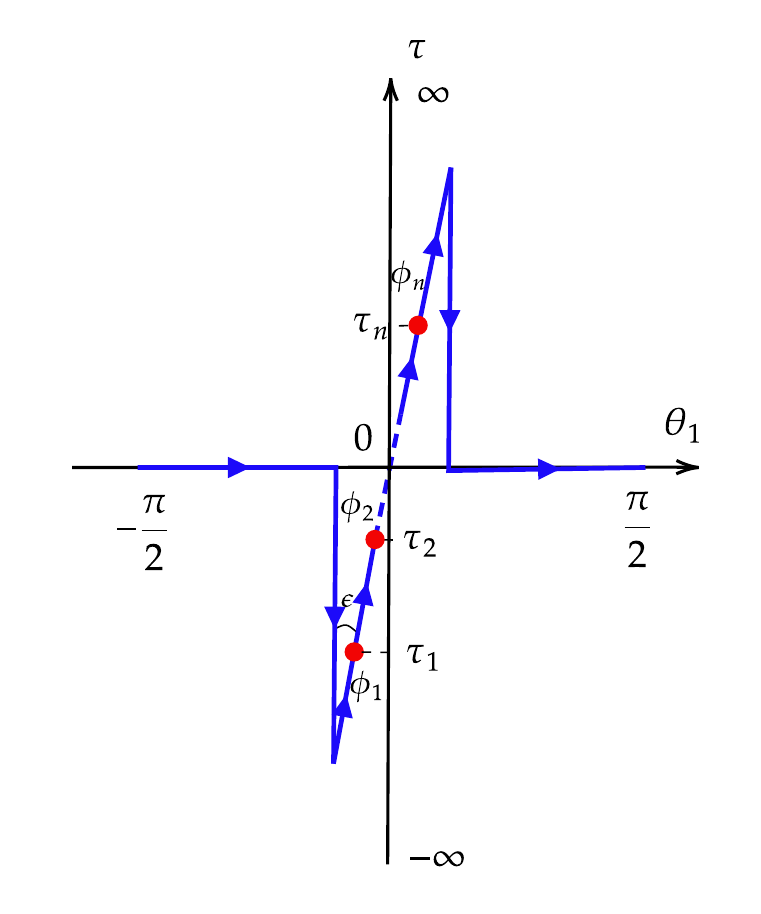}
    } \quad
    \subfigure[The $\theta_1$-ordered $n$-point correlator $\bra{0} \phi_n (x_n) \cdots \phi_1 (x_1) \ket{0}$, where $\theta_1^1 < \cdots < \theta_1^n$. \label{fig:dSSncorrelator}]{
    \includegraphics[width=0.45\linewidth]{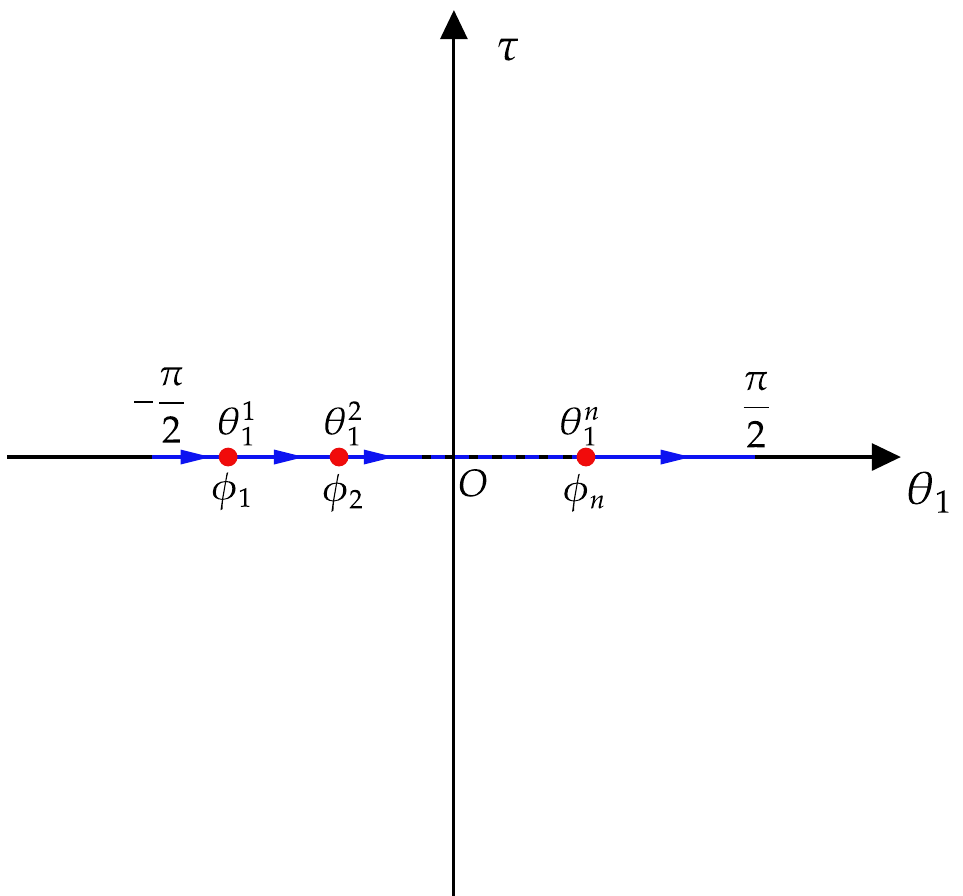}
    }
    \caption{The path integral approach to calculate the correlators in dS$_n$ and $S^n$. The complexified $\theta_1$-plane used to describe both dS$_n$ and $S^n$ is parameterized by $\theta_1 = i \tau$.}
\end{figure}

The Euclidean vacuum $|0\>$ is defined by a Euclidean path integral over the lower hemisphere $\theta_1\leq 0$ 
\begin{equation}
    \<\tilde{\phi}|0\>= \int_{\phi(\theta_1=0^-)=\tilde{\phi}} D\phi \exp\left(\int_{-\frac{\pi}{2}}^{0^-}\td\theta_1\int\td\Omega_{n-1}\sqrt{g} \mathcal{L}(\phi)\right)\,,
\end{equation}
while the conjugation bra $\<0|$ is defined by the path integral over the upper hemisphere $\theta_1\geq0$. 
\begin{equation}
    \<0|\tilde{\phi}\>= \int_{\phi(\theta_1=0^+)=\tilde{\phi}} D\phi \exp\left(\int_{0^+}^{\frac{\pi}{2}}\td\theta_1\int\td\Omega_{n-1}\sqrt{g} \mathcal{L}(\phi)\right)\,.
\end{equation}
The vacuum partition function is given by
\begin{equation}\label{eq:parti-dS}
    Z_{\text{dS}_n} [0] = \<0|0\>= \int D\phi \exp\left(\int_{S^n}\sqrt{g} \td^n x \mathcal{L}(\phi)  \right)\,.
\end{equation}
The Feynman correlation functions in the Euclidean vacuum $\<0|T\phi(x_1)\cdots\phi(x_n)|0\>$, where $T$ is the ``time''-ordering operator placing operators with larger (smaller) $\tau$ to the left (right), are given by the path integral over the region depicted in blue, as shown in Fig.~\ref{fig:pathCorretheta}. The field insertions are depicted by red dots. The path integral must pass through these red dots in the $\tau$‑ordered sequence, which prevents the integration contour from being deformed onto the real axis, since the locations of the red dots are fixed. Effectively, the range of the slightly tilted blue line in Fig.~\ref{fig:pathCorretheta} only needs to cover all the red dots and does not need to extend to infinity. However, in that case, the integration region would depend on the specific correlator. To avoid this dependence, it is more convenient to choose the slightly tilted blue line extending from $-i\infty(1 - i\epsilon)$ to $i\infty(1 - i\epsilon)$, which provides a common integration region for all correlators. 

Importantly, since the blue segment on the real axis in Fig.~\ref{fig:pathCorretheta} is introduced to pick out the BD vacuum in the dS space, its position cannot be deformed either, unless one could perform a Wick rotation that moves it onto the slightly tilted blue line. However, such a rotation is not allowed here, because the upper bound of $\theta_1$ is finite at $\pi/2$. Under this choice, we denote the entire blue region in Fig.~\ref{fig:pathCorretheta} by $\mathcal{C}$. Therefore, the sourced partition function is defined by inserting a sourced vertex operator into the region $\mathcal{C}$
\begin{equation} \label{eq:ZdSJ}
    Z_{\text{dS}_n} [\mathcal{J}] = \int D\phi \exp\left[\int_{\mathcal{C}}\sqrt{g} \td^n x (\mathcal{L}(\phi) + \mathcal{J} (x) \phi(x)) \right]\,.
\end{equation}
Here, the $\sqrt{g} \td^n x = \cos^{n-1} \theta_1 \td \theta_1 \td^{n-1} \Omega$ is formally the ``volume element'' of the $S^n$ metric \eqref{eq:Sn} but with complex $\theta_1$. The operation of taking functional derivatives with respect to $\mathcal{J} (x_i)$ with $\theta_1^i \in(-i\infty(1 - i\epsilon), i\infty(1 - i\epsilon))$ is equivalent to inserting fields $\phi(x_i)$ in this region. In particular, if there are no field insertions ($\mathcal{J}=0$), the integral region can be smoothly deformed to the real axis integrating $\theta_1$ from $-\pi/2$ to $\pi/2$, which returns to the definition of the vacuum partition function \eqref{eq:parti-dS}.

The canonical quantization in dS$_n$ with the metric \eqref{eq:metricdS} is performed on the hypersurface $\Sigma_\tau: \tau =\text{Constant}$, corresponding to the canonical quantization on the hypersurface $\Sigma_{\theta_1}: \theta_1 =\text{Constant}$ in $S^n$.

Similar to the treatment in \cite{Chernikov:1968zm}, we rewrite the metric of the sphere $S^n$ \eqref{eq:Sn} in a more convenient way with $\tanh \eta = \sin \theta_1$
\begin{equation} \label{eq:Sneta}
    \td \Omega_n^2 = \td s^2/l^2 = (\cosh \eta)^{-2} (\td \eta^2 + \td \Omega_{n-1}^2), \quad \eta \in (-\infty, \infty),
\end{equation}
Under this coordinate system, the hypersurface $\Sigma_{\theta_1}$ is equivalent to $\Sigma_\eta$. Moreover, the vacua $\bra{0}$ and $\ket{0}$ are defined via the Euclidean path integral by integrating over the regions $\eta \in (0, \infty)$ (or $\theta_1 \in (0, \pi/2)$) and $\eta \in (- \infty, 0)$ (or $\theta_1 \in (- \pi/2, 0)$), respectively. Consequently, the convergence of the Euclidean path integral requires the following regularity condition
\begin{equation} \label{eq:regulardS}
    \bra{0} \phi (\eta \rightarrow \infty, \vec \theta), ~ \phi (\eta \rightarrow -\infty, \vec \theta) \ket{0} \rightarrow \text{regular}.
\end{equation}
Here we do not distinguish the notation of Euclidean vacua in dS$_n$ and $S^n$, as they share the same integral regions on the complex plane.

Using the coordinates in \eqref{eq:Sneta}, the action for a free scalar field in $S^n$ can be written as
\begin{equation}
    \begin{split}
         S & = - \frac12 \int \td^n x \sqrt{g} \left[\nabla_\mu \phi \nabla^\mu \phi + m^2 \phi^2 \right] \\
         & = -\frac{l^{n-2}}{2} \int^{\infty}_{-\infty} \td \eta (\cosh \eta)^{2-n} \int \td^{n-1} \Omega \left[(\p_\eta \phi)^2 + \p_A \phi \p^A \phi + \frac{m^2 l^2 \phi^2}{(\cosh \eta)^2} \right],
    \end{split}
\end{equation}
where $\td^{n-1} \Omega = \sqrt{h^{S^{n-1}}} \td^{n-1} \theta ~ (\td^{n-1} \theta = \td \theta_2 \cdots \td \theta_n)$ and $h^{S^{n-1}}_{AB}$ are the volume element and the metric of $S^{n-1}$. Here, we also used the abbreviation $x^A = (\theta_2, \cdots, \theta_n)$, and the index ``$A$'' is raised and lowered by the metric of $S^{n-1}$. The conjugate momentum is then
\begin{equation}
    \pi_\eta = \frac{\delta S}{\delta (\p_\eta \phi)} = -l^{n-2} (\cosh \eta)^{2-n} \sqrt{h^{S^{n-1}}} \p_\eta \phi.
\end{equation}
Then, the canonical quantization yields
\begin{equation} \label{eq:dScano}
    [\phi (\eta, \vec \theta), \pi_\eta (\eta, \vec \theta')] = -\delta^{n-1} (\vec \theta - \vec \theta'), \quad [\phi (\eta, \vec \theta), \phi (\eta, \vec \theta')] = 0.
\end{equation}
We also need the definition of Klein inner product
\begin{equation} \label{eq:KleininndS}
    \<\phi^1, \phi^2\> = - l^{2-n} \int_{\Sigma_\eta} \td^{n-1} \theta (\phi^1 \pi^2_\eta - \phi^2 \pi_\eta^1) = (\cosh \eta)^{2-n} \int_{S^{n-1}} \td^{n-1} \Omega \left[ \phi^1 \overset{\leftrightarrow}{\p_\eta} \phi^2 \right]
\end{equation}

The equation of motion (Klein-Gordon equation) is
\begin{equation}
    l^2(\nabla^2 - m^2) \phi = [(\cosh \eta)^n \p_\eta ((\cosh \eta)^{2-n} \p_\eta) + (\cosh \eta)^2 \Delta_{n-1} - m^2 l^2] \phi = 0,
\end{equation}
where $\Delta_{n-1}$ is the Laplace operator on $S^{n-1}$, whose eigenfunctions are the hyperspherical harmonics $Y^{(n)}_{\ell, \{\mu\}}$ \cite{wen1985some}
\begin{equation}
    \Delta_{n-1} Y^{(n)}_{\ell, \{\mu\}} (\vec \theta) = - \ell (\ell + n -2) Y^{(n)}_{\ell, \{\mu\}} (\vec \theta),
\end{equation}
which form an orthonormal and complete basis,
\begin{equation} \label{eq:orthcomp}
    \begin{split}
        & \int \td^{n-1} \Omega Y^{(n)}_{\ell, \{\mu\}} (\vec \theta) Y^{(n)}_{\ell', \{\mu'\}} (\vec \theta) = \delta_{\ell, \ell'} \delta_{\{\mu\}, \{\mu'\}}\,; \\
        & \sum_{\ell, \{\mu\}} Y^{(n)}_{\ell, \{\mu\}} (\vec \theta) Y^{(n)}_{\ell, \{\mu\}} (\vec \theta') = \frac{1}{\sqrt{h^{S^{n-1}}}} \delta^{n-1} (\vec \theta - \vec \theta').
    \end{split}
\end{equation}
Here we choose the hyperspherical harmonics to be real, and we used the abbreviation $\{\mu\} = \{\mu_1, \cdots, \mu_{n-2}\}$. The range of the indices in the summation is $\ell \geq \mu_1 \geq \cdots \geq \mu_{n-2} \geq 0$.

The scalar field $\phi$ can be expanded through $Y^{(n)}_{s, \{\mu\}}$ as
\begin{equation}
    \phi (x) = \sum_{\ell, \{\mu\}} Y^{(n)}_{\ell, \{\mu\}} (\vec \theta) u (\eta) a_{\ell, \{\mu\}}.
\end{equation}
Define
\begin{equation}
    u(\eta)=(\cosh\eta)^{\frac{n-2}{2}}v(\eta)\,,
\end{equation}
the equation of motion becomes
\begin{equation} \label{eq:udS}
    v''(\eta) - \left(p^2 + \frac{\bar M (1- \bar M)}{(\cosh \eta)^2} \right) v (\eta) = 0
\end{equation}
where
\begin{equation}
    p = \ell + \frac{n-2}{2}, \quad \bar M = \frac12 - \sqrt{\frac{(n-1)^2}{4} - m^2 l^2}\,.
\end{equation}

The equation \eqref{eq:udS} has two independent solutions, and the appropriate choice can be 
\begin{equation}
    \begin{split}
        u_p^\pm (\eta) = & 2^{\bar M} \frac{\sqrt{\Gamma(p+ \bar M) \Gamma(p- \bar M +1)}}{\Gamma(p+1)} (\cosh \eta)^{\bar M + \frac n2-1} e^{\pm (p + \bar M) \eta} \\
        & \times {_2F_1} \left( p+\bar M, \bar M;p+1;-e^{\pm 2 \eta} \right),
    \end{split}
\end{equation}
which are linearly independent, as their Wronskian, defined by the $\eta$-part in the Klein inner product \eqref{eq:KleininndS}, is nonzero
\begin{equation}
    W [u^-_p, u^+_p] = (\cosh \eta)^{2-n} \left[ u^-_p (\eta) \overset{\leftrightarrow}{\p_\eta} u^+_p (\eta) \right] = 2.
\end{equation}
One can find a similar property of the corresponding modes in dS$_n$ space \cite{Chernikov:1968zm}. This choice of the modes $u^\pm_p$ is appropriate for defining the vacuum unambiguously, as their asymptotic behavior, presented in Table \ref{tab:dSmode}, does not diverge simultaneously at the two boundaries $\eta \rightarrow \pm \infty$. As a result, with the mode expansion of the field
\begin{equation} \label{eq:phidS}
    \phi (\eta, \vec \theta) = \sum_{\ell, \{\mu\}} Y^{(n)}_{\ell, \{\mu\}} (\vec \theta) \left[ u_p^+ (\eta) a^+_{\ell, \{\mu\}} + u_p^- (\eta) a^-_{\ell, \{\mu\}} \right],
\end{equation}
one can explicitly express the regularity condition \eqref{eq:regulardS} as
\begin{equation}
    \bra{0} a^+_{\ell, \{\mu\}} = 0, \qquad a^-_{\ell, \{\mu\}} \ket{0} = 0.
\end{equation}

\begin{table}[htbp]
    \centering
    \begin{tabular}{c|c|c}
         \hline
         $\eta$ & $u^+_p (\eta)$ & $u^-_p (\eta)$ \\
         \hline
         $-\infty$ & $0 (\ell \neq 0)$ or $\frac{\sqrt{\Gamma(p+ \bar M) \Gamma(p- \bar M +1)}}{2^{(n-2)/2}\Gamma(p+1)} (\ell=0)$ & $\infty$ \\
         \hline
         $+\infty$ & $\infty$ & $0 (\ell \neq 0)$ or $\frac{\sqrt{\Gamma(p+ \bar M) \Gamma(p- \bar M +1)}}{2^{(n-2)/2} \Gamma(p+1)} (\ell=0)$\\
         \hline
    \end{tabular}
    \caption{The asymptotic behavior of the modes $u^\pm_p$.}
    \label{tab:dSmode}
\end{table}

Moreover, by expressing the operators $a^\pm_{\ell, \{\mu\}}$ using the Klein inner product between the modes and the field,
\begin{equation}
    \begin{split}
        a^\pm_{\ell, \{\mu\}} = & \frac{\pm1}{W [u^-_p, u^+_p]} \left\< Y^{(n)}_{\ell, \{\mu\}} (\vec \theta) u_p^\mp (\eta), \phi(\eta, \vec \theta) \right\> \\
        = & \pm \frac12 (\cosh \eta)^{2-n} \int_{S^{n-1}} \td^{n-1} \Omega Y^{(n)}_{\ell, \{\mu\}} (\vec \theta) \left[ u_p^\mp (\eta) \overset{\leftrightarrow}{\p_\eta} \phi(\eta, \vec \theta) \right]\,,
    \end{split}
\end{equation}
the canonical commutators can be re-expressed as
\begin{equation}
    \left[ a^-_{\ell, \{\mu\}}, a^+_{\ell', \{\mu'\}} \right] = \frac{l^{2-n}}{2} \delta_{\ell, \ell'} \delta_{\{\mu\}, \{\mu'\}}.
\end{equation}

Finally, the $\eta$-ordered propagator is then derived as
\begin{equation} \label{eq:Etacorrelator}
    \begin{split}
        \bra{0} E \phi (x) \phi (x') \ket{0} & = \Theta(\eta - \eta') \bra{0} \phi (x) \phi (x') \ket{0} + \Theta(\eta' - \eta) \bra{0} \phi (x') \phi (x) \ket{0} \\
        &= \bra{0} 0 \> \frac{\Gamma(\frac{n-2}{2})}{4 \pi^{n/2} l^{n-2}} \sum_{\ell = 0}^\infty p C_\ell^{(\frac{n-2}{2})} (\vec \theta \cdot \vec \theta') u^+_p (\eta_<) u^-_p (\eta_>)
    \end{split}
\end{equation}
where $\Theta(\eta)$ is a step function equal to 1 when $\eta>0$ and equal to 0 when $\eta <0$, $E$ denotes the $\eta$-ordered operator that places the field with larger $\eta$ on the left, $x_< (x_>)$ represents the coordinates with smaller (larger) $\eta$ among $x, x'$, and the Gegenbauer polynomial is given by \cite{wen1985some}
\begin{equation} \label{eq:Geg}
    C_\ell^{(\frac{n-2}{2})} (\vec \theta \cdot \vec \theta') = \frac{2 \pi^{\frac{n}{2}}}{p \Gamma(\frac{n}{2}-1)} \sum_{\{ \mu\}} Y^{(n)}_{\ell, \{\mu\}} (\vec \theta) Y^{(n)}_{\ell, \{\mu\}} (\vec \theta')\,.
\end{equation}

The $\eta$-ordered correlator satisfies the Dyson-Schwinger equation
\begin{equation}
    (-\nabla_x^2 + m^2) G_E (x-x') = \frac{1}{\sqrt{g}} \delta^n (x-x'), \quad G_E (x-x') = \frac{\bra{0} E \phi (x) \phi (x') \ket{0}}{\bra{0} 0 \> },
\end{equation}
which can be verified by using $\p_\eta \Theta (\eta - \eta') = \delta (\eta - \eta')$, the equation of motion of the field $(\nabla^2 - m^2) \phi =0$, and the canonical commutators \eqref{eq:dScano}. Moreover, since the propagator \eqref{eq:Etacorrelator} is SO$(n+1)$ invariant, it only depends on the geodesic distance $\sigma$ between $x, x'$ in $S^n$, where
\begin{equation}
    \cos \sigma = \frac{\cos \Omega_{n-1}}{\cosh \eta \cosh \eta'} + \tanh \eta \tanh \eta'
\end{equation}
with $\Omega_{n-1}$ representing the geodesic distance between $\vec \theta, \vec \theta'$ on a unit $S^{n-1}$.

In particular, one can always choose $\eta_> = - \eta_< = \eta_+ (>0)$ by an SO$(n+1)$ transformation. The propagator can then be parameterized as $G_E (x-x') = G_E (\eta_+, \cos \Omega_{n-1})$. Furthermore, by using the results in Table \ref{tab:dSmode}, the asymptotic behavior of $G_E$ at $\eta_+ \rightarrow \infty$ (or equivalently, $\cos \sigma \rightarrow -1$) can be derived as
\begin{equation}
    G_E (\eta_+ \rightarrow \infty, \cos \Omega_{n-1}) \rightarrow \frac{\Gamma(\delta_+) \Gamma(\delta_-)}{\Gamma (\frac{n}{2}) (4\pi)^{n/2} l^{n-2}},
\end{equation}
where we also used the property of the Gegenbauer polynomial $C_0^\alpha (x) \equiv 1$. Therefore, the series \eqref{eq:Etacorrelator} can actually be summed in a compact form
\begin{equation}
    \bra{0} E \phi (x) \phi (x') \ket{0}/ \bra{0} 0\> = \frac{\Gamma(\delta_+)\Gamma(\delta_-)}{l^{n-2}2^n\pi^{n/2}\Gamma(n/2)} {_2F_1} \left(\delta_-,\delta_+;\frac{n}{2};\frac{1+\cos\sigma}{2}\right)\,.
\end{equation}
The right-hand side also satisfies the Dyson-Schwinger equation and has the same asymptotic behavior at $\cos \sigma \rightarrow -1$. This can be easily checked by using the property of the hypergeometric function ${_2F_1} (a,b;c;z=0) \equiv 1$. This result agrees exactly with the two-point Wightman function \eqref{eq:SnCorrelator} on $S^n$.

The $\eta$-ordered correlation function in $S^n$ can be described by a path integral over the blue line as illustrated by Fig.~\ref{fig:dSSncorrelator}. Unlike the $\tau$-ordered dS$_n$ correlators depicted in Fig.~\ref{fig:pathCorretheta}, the fields in $S^n$ are inserted on the real axis, which can be mapped to the imaginary axis with the substitution $\theta_1 \rightarrow i \tau (1-i \epsilon)$ (or equivalently, $\eta \rightarrow i \theta (1-i \epsilon)$) for the fields. Moreover, after this substitution, the above results all return to the data derived by canonical quantization on the hypersurface $\Sigma_\tau$ in dS$_n$ space, including the mode expansion and correlation function, as discussed in \cite{Mottola:1984ar, Bousso:2001mw}.

\section{Two-point function of flipped CFT$_{n-1}$} \label{sec:flipCFT}

The flipped CFT$_{n-1}$ is defined on a Lorentzian torus, denoted by $T^{n-2,1}=S^{n-2}\times S^{1}$ with metric
\begin{equation}
    \td s^2 \propto \td t^2-\td\Omega_{n-2}^2\,,\quad t\sim t+2\pi\,.
\end{equation}
It has a conformal symmetry SO($n-1,2$), which is the same as the isometry group of the flipped AdS$_n/\mathbb{Z}$ space.

The CFT living in spacetime with the metric $\td s^2\propto \td t^2-\td\Omega_{n-2}^2$ differs from that with the metric $\td s^2\propto -\td t^2+\td\Omega_{n-2}^2$ by a flipping transformation $g_{\mu\nu}\to g_{\mu\nu}'=-g_{\mu\nu}$, which flips the signature of the metric, or equivalently interchanges the timelike and spacelike directions. Note that the conformal Ward identity is the same for the case of $T^{n-2,1}$ and the case of $T^{1,n-2}$. Although the conformal Ward identities are identical for $T^{n-2,1}$ and $T^{1,n-2}$, their correlation functions differ through an overall phase and the sign of the Feynman $i\epsilon$-prescription, reflecting their distinct causal structures.

First, we consider a standard CFT$_{n-1}$ living in a Euclidean plane $\mathbb{R}^{n-1}$. The correlation function of two primary operators with conformal dimension $\Delta$ is
\begin{equation}
    \<O(x)O(x')\>_{\mathbb{R}^{n-1}}=\frac{1}{(x-x')^{2\Delta}}\,.
\end{equation}
We rewrite the plane coordinates as $x^i=e^{t_E} y^i$ with $y^i$ being a unit vector satisfying $y\cdot y=1$. Then the metric becomes
\begin{equation}
    \td s^2_{\text{plane}}=e^{2t_E}(\td t_E^2+\td\Omega_{n-2}^2)\,.
\end{equation}
We perform a Weyl transformation to obtain a CFT correlation function in a Euclidean cylinder,
\begin{equation}
    \td s^2_{\text{cyl}}=\td t_E^2+\td\Omega_{n-2}^2\,.
\end{equation}
The correlation function becomes
\begin{equation}
    \<O(t_E,\Omega_{n-2})O(t_E',\Omega_{n-2}')\>_{\text{cyl}}=\left|\frac{\partial x}{\partial(t_E,\Omega_{n-2})}\right|^{\Delta/(n-1)}\left|\frac{\partial x'}{\partial(t_E',\Omega_{n-2}')}\right|^{\Delta/(n-1)}\<O(x)O(x')\>_{\text{plane}}\,,
\end{equation}
where the Jacobian determinant is
\begin{equation}
    \left|\frac{\partial x}{\partial(t_E,\Omega_{n-2})}\right|=e^{(n-1)t_E}\,.
\end{equation}
The final result is
\begin{equation}\label{eq:EuclideanCylinder1}
    \<O(t_E,\Omega_{n-2})O(t_E',\Omega_{n-2}')\>_{\text{cyl}}=\frac{1}{\left(2\cosh (t_E-t_E')-2 y\cdot y'\right)^\Delta}\,.
\end{equation}
This result can also be expressed as
\begin{equation}\label{eq:EuclideanCylinder2}
    \<O(t_E,\Omega_{n-2})O(t_E',\Omega_{n-2}')\>_{\text{cyl}}=\frac{1}{\left(2\sin \frac{\delta\Omega-i(t_E-t_E')}{2}\right)^\Delta\left(2\sin \frac{\delta\Omega+i(t_E-t_E')}{2}\right)^\Delta}\,,
\end{equation}
where $\delta \Omega=\arccos(y\cdot y')\in[0,\pi]$.

\begin{figure}[htbp]
    \centering
    \subfigure[The poles of the two-point correlation function $\<O(t,\Omega_{n-2})O(t',\Omega_{n-2}')\>_{\text{cyl}}$ given by Eq.~\eqref{eq:2ptCylinderCos} or Eq.~\eqref{eq:2ptCylinderSin}. \label{fig:Wickploes1}]{
    \includegraphics[width=0.8\linewidth]{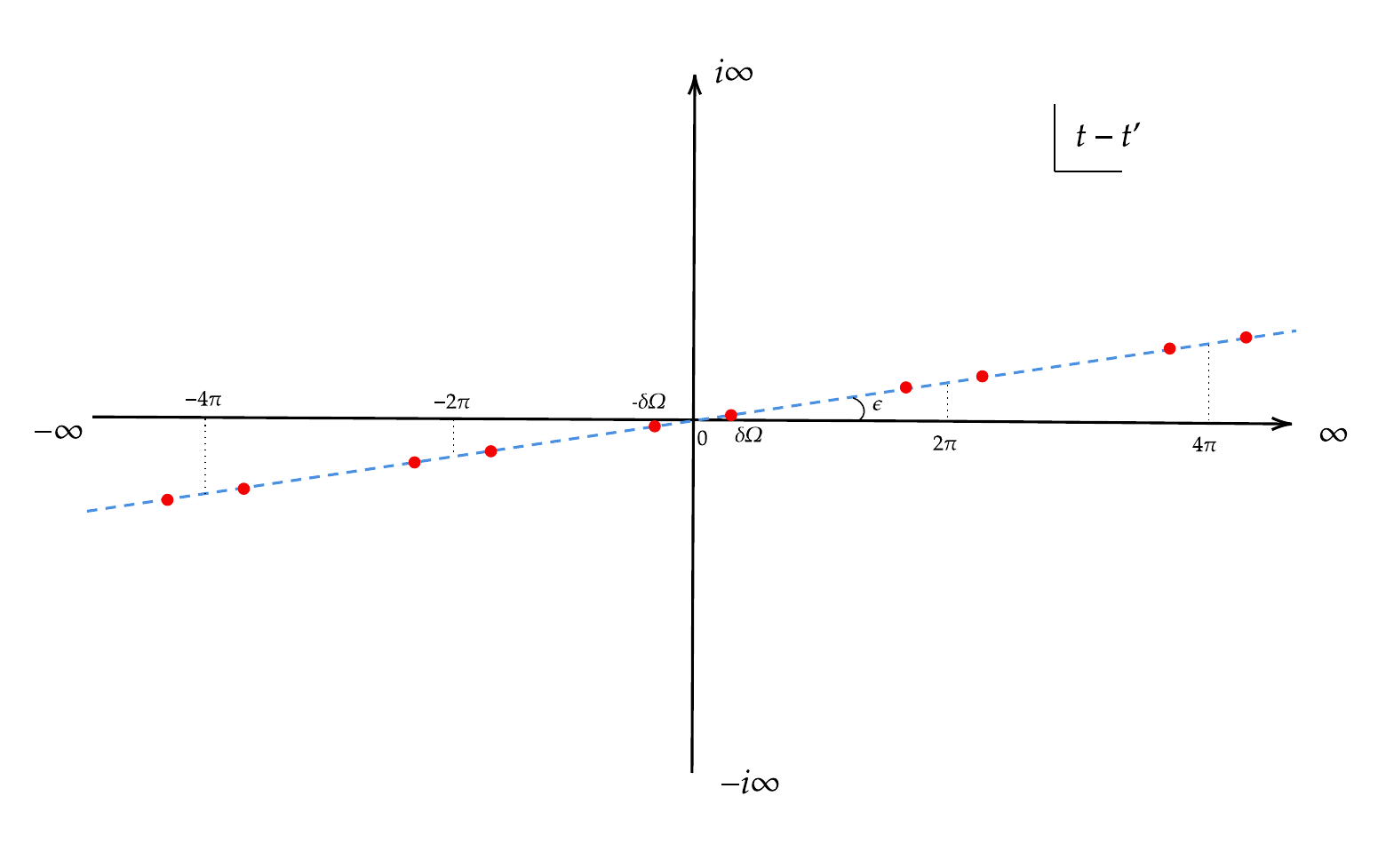}
    } \\
    \subfigure[The poles of the two-point correlation function $\<O(t,\Omega_{n-2})O(t',\Omega_{n-2}')\>_{T^{1,n-2}}$ given by Eq.~\eqref{eq:CFT2ptT^1,n-2} or Eq.~\eqref{eq:CFT2ptT^1,n-2Sin}. \label{fig:Wickpoles2}]{
    \includegraphics[width=0.8\linewidth]{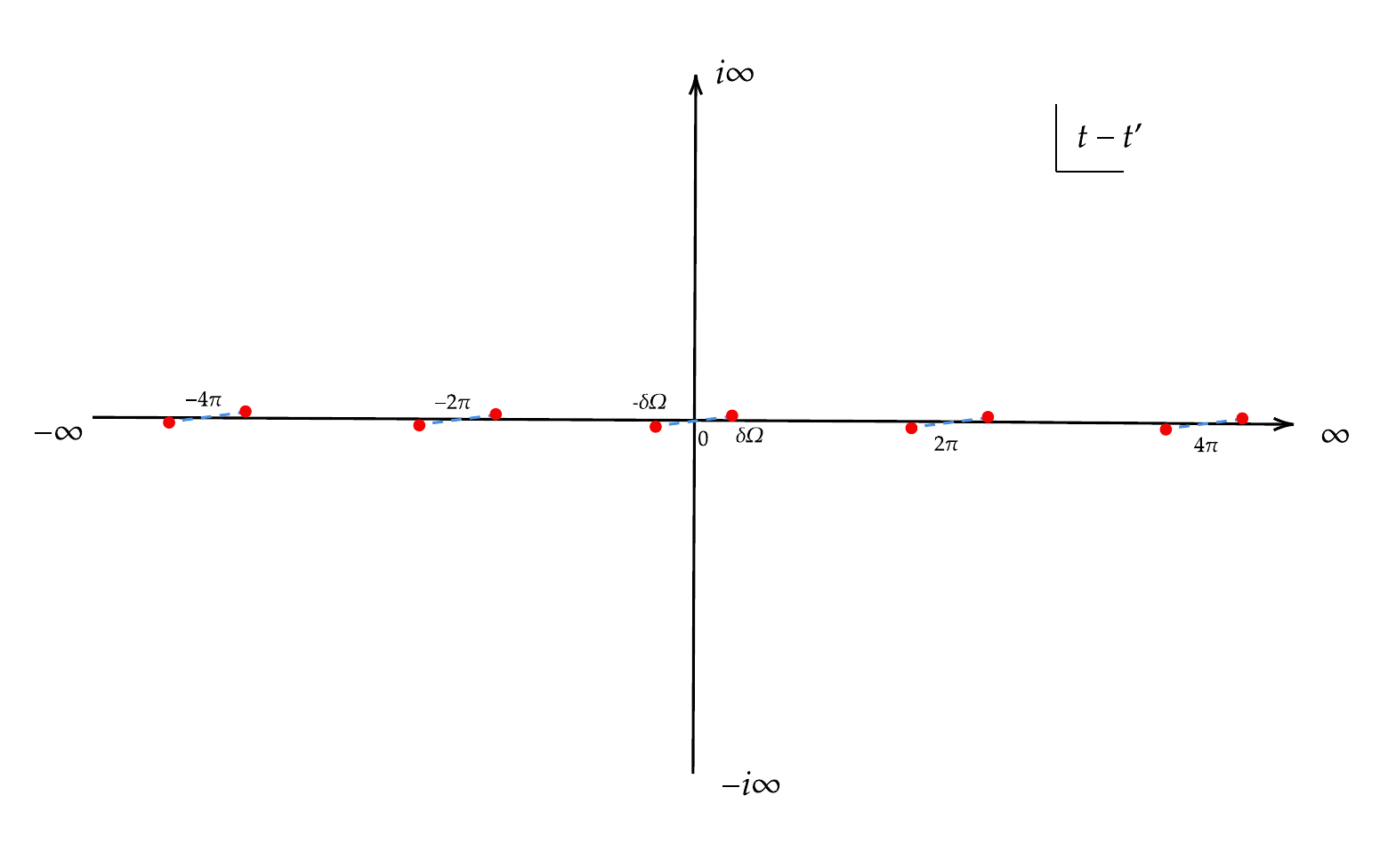}
    }
    \caption{The poles of the correlation functions are depicted with red dots. For real-time $t$, whether the poles appear on the upper half-plane or the lower half-plane determines how the integration contour bypasses the light cone poles.}
\end{figure}

From the result in the Euclidean cylinder, we can obtain the correlation function in a Lorentzian cylinder,
\begin{equation}
    \td s^2=-\td t^2+\td\Omega_{n-2}^2\,,\quad t\in(-\infty,\infty)\,,
\end{equation}
by the analytic continuation $t_E=it(1-i\epsilon)$. The result \eqref{eq:EuclideanCylinder1} becomes
\begin{equation}\label{eq:2ptCylinderCos}
    \<O(t,\Omega_{n-2})O(t',\Omega_{n-2}')\>_{\text{cyl}}=\frac{1}{\left(2\cos((t-t')(1-i\epsilon))-2 y\cdot y'\right)^\Delta}\,,
\end{equation}
and the result \eqref{eq:EuclideanCylinder2} becomes
\begin{equation}\label{eq:2ptCylinderSin}
    \<O(t,\Omega_{n-2})O(t',\Omega_{n-2}')\>_{\text{cyl}}=\frac{1}{\left(2\sin \frac{\delta\Omega+(t-t')(1-i\epsilon)}{2}\right)^\Delta\left(2\sin \frac{\delta\Omega-(t-t')(1-i\epsilon)}{2}\right)^\Delta}\,,
\end{equation}
These two expressions are equivalent to each other with the $i\epsilon$-prescription introduced in the Wick rotation $t_E=i t(1-i\epsilon)$. They seem to be periodic under $t\to t+2\pi$ since the variable $(t-t')$ appears in the trigonometric functions. However, they acquire an additional phase under such discrete translations. This can be seen by the position of the poles associated with the light cone,
\begin{equation}\label{eq:poleCylinder}
    t-t'=(\pm\delta\Omega+2n\pi)(1+i\epsilon)\,,\quad n=0,\pm1,\pm2\cdots
\end{equation}
They are not periodic under $t\to t+2\pi$ as shown in Fig.~\ref{fig:Wickploes1}. However, the holographic two-point function \eqref{eq:powerlawBranch} is periodic. Fortunately, we can choose another $\epsilon$-prescription since it cannot be determined through symmetries. We may instead replace the prescription in Eq.~\eqref{eq:poleCylinder} by
\begin{equation}\label{eq:poleTorus}
    t-t'=\pm\delta\Omega(1+i\epsilon)+2n\pi\,,
\end{equation}
In this case, the correlators are now periodic under $t \to t + 2\pi$, as shown in Fig.~\ref{fig:Wickpoles2}. The resulting two-point function is
\begin{equation}\label{eq:CFT2ptT^1,n-2}
    \<O(t,\Omega_{n-2})O(t',\Omega_{n-2}')\>_{T^{1,n-2}}=\frac{1}{\left(2\cos(t-t')-2 y\cdot y'+i\epsilon\right)^\Delta}\,,
\end{equation}
or equivalently
\begin{equation}\label{eq:CFT2ptT^1,n-2Sin}
    \<O(t,\Omega_{n-2})O(t',\Omega_{n-2}')\>_{T^{1,n-2}}=\frac{1}{\left(2\sin \frac{\delta\Omega+(t-t')+i\epsilon}{2}\right)^\Delta\left(2\sin \frac{\delta\Omega-(t-t')+i\epsilon}{2}\right)^\Delta}\,.
\end{equation}

Recall that we have obtained the correlation function \eqref{eq:powerlawBranch} in the Lorentzian torus $T^{n-2,1}$ by holography. Comparing the correlator in Eq.~\eqref{eq:CFT2ptT^1,n-2} for $T^{1,n-2}$ with that in Eq.~\eqref{eq:powerlawBranch} for $T^{n-2,1}$, we find that they can be identified by taking $\epsilon\to -\epsilon/2, c_{\pm}=1/(-2)^\Delta$ and $\delta_{\pm}=\Delta$ in Eq.~\eqref{eq:powerlawBranch}. Thus, from the perspective of the correlation functions, the flipped CFT$_{n-1}$ has a structure similar to that of a standard CFT$_{n-1}$ but with a different causal structure.

The CFT correlation function in a Lorentzian cylinder can be interpreted as a correlation function in a Lorentzian torus ($t\sim t+2\pi$) because the two geometries share the same conformal symmetry $\mathrm{SO}(n-1,2)$. This reflects the fact that the AdS$/\mathbb{Z}$ spacetime has the isometries SO($n-1,2$) if and only if $t\sim t+2\pi$. We see that the conformal group is broken to SO$(n-1)\times$SO$(2)$ when the period of the coordinate $t$ is not exactly $2\pi$ in Section \ref{sec:kerrdS}.

\section{Asymptotic fAdS$_3$ and modular parameter} \label{sec:conifoldS3}

We start from the embedding description of a Euclidean sphere $S^3$
\begin{equation} \label{eq:S3embed}
    \td s^2 = (\td x^1)^2+(\td x^2)^2+(\td x^3)^2+(\td x^4)^2, \quad (x^1)^2+(x^2)^2+(x^3)^2+(x^4)^2 = l^2.
\end{equation}
We first parameterize the patch of $S^3$ relevant for the Kerr-dS static region by
\begin{equation} \label{eq:S3dS}
    \begin{split}
        x^1 & = l \sqrt{\frac{r_+^2 - r^2}{r_+^2 + r_-^2}} \sin \chi, \quad x^2 = l \sqrt{\frac{r_+^2 - r^2}{r_+^2 + r_-^2}} \cos \chi, \\
        x^3 & = l \sqrt{\frac{r^2 + r_-^2}{r_+^2 + r_-^2}} \sin \xi, \quad x^4 = l \sqrt{\frac{r^2 + r_-^2}{r_+^2 + r_-^2}} \cos \xi.
    \end{split}
\end{equation}
Then, the metric of $S^3$ can be rewritten as
\begin{equation}
    \td s^2 /l^2 = \td\Omega_3^2 = \frac{r^2 \td r^2}{(r_+^2 - r^2) (r^2 + r_-^2)} + \frac{r^2 + r_-^2}{r_+^2 + r_-^2} \td \xi
    ^2 + \frac{r_+^2 - r^2}{r_+^2 + r_-^2} \td \chi^2,
\end{equation}
where $r<r_+; ~ \chi, \xi \in [0, 2\pi)$, and $r_+^2, -r_-^2$ are the solutions of $\mathcal{M} - \frac{r^2}{l^2} + \frac{J^2}{4r^2} = 0$:
\begin{equation} \label{eq:r+-}
    r_\pm^2 = \frac{1}{2} \left[ \sqrt{(\mathcal{M} l^2)^2 + J^2 l^2} \pm \mathcal{M} l^2 \right].
\end{equation}

As demonstrated in Section \ref{sec:embed}, the complexification $\chi \rightarrow i \chi_*$ in \eqref{eq:S3dS} is equivalent to $x^1 \rightarrow i t$ in the embedding \eqref{eq:S3embed}, which gives the definition of dS$_3$. Then, together with the transformation
\begin{equation} \label{eq:xitautrans}
    \xi = \frac{r_+}{l} \phi - \frac{r_-}{l^2} t_*, \quad \chi_* = \frac{r_-}{l} \phi + \frac{r_+}{l^2} t_*,
\end{equation}
one obtains
\begin{equation} \label{eq:KerrdS}
    \td s^2 = - N^2 \td t_*^2 + N^{-2} \td r^2 + r^2 (\td \phi - N^\phi \td t_*)^2,
\end{equation}
where
\begin{equation}
    N^2 = \frac{(r_+^2-r^2) (r^2 + r_-^2)}{l^2 r^2} = \mathcal{M} - \frac{r^2}{l^2} + \frac{J^2}{4r^2}, \quad N^\phi = \frac{r_+ r_-}{l r^2} = \frac{J}{2 r^2}.
\end{equation}
This is the metric of Kerr-dS$_3$ \cite{Park:1998qk,Balasubramanian:2001nb}, which is locally equivalent to dS$_3$.

In the embedding description, the periodicity $2 \pi$ of the angular coordinate $\chi$ in the Euclidean space $\mathbb R^4$ ensures the absence of a conical singularity of the embedded hypersurface (in this case, Kerr-dS$_3$). This, in turn, implies that the time direction $\chi_* = - i \chi$ acquires an imaginary period $-2 \pi i$, giving rise to thermal effects on the hypersurface. Correspondingly, the periodicity caused by the thermal effect can be expressed as 
\begin{equation} \label{eq:thermalchi}
    (\xi, \chi_*) \simeq (\xi, \chi_* - 2 \pi i)
\end{equation}
or, equivalently, can be expressed in terms of $(\phi, t_*)$ in \eqref{eq:xitautrans} as
\begin{equation}
    \frac{t_*}{l} \simeq \frac{t_*}{l} - \frac{2 \pi i l r_+}{r_+^2 + r_-^2}, \qquad \phi \simeq \phi - \frac{2 \pi i l r_-}{r_+^2 + r_-^2}.
\end{equation}
It can be easily checked that the metric \eqref{eq:KerrdS} and this thermal periodic condition are equivalent to \eqref{eq:S3dSconifold} and \eqref{eq:periodKerrdS} after the Wick rotation $t_*=-i t$ and the rescaling \eqref{eq:rescaling}.

Furthermore, the spatial period of $\phi \simeq \phi + 2 \pi$ induces additional identifications of the coordinates $(\xi, \theta)$ of $S^3$, 
\begin{equation} \label{eq:Gamma+}
    \Gamma_\phi:~ (\xi, \chi) \simeq \left(\xi + 2 \pi \frac{r_+}{l}, \chi + 2 \pi i \frac{r_-}{l} \right).
\end{equation}
This causes the conical defect on $S^3$, implying that Kerr-dS$_3$ is obtained by analytic continuation from the conifold $S^3/\Gamma_\phi$.

Then, we provide an alternative analytic continuation of the conifold $S^3/\Gamma_\phi$ to a spacetime that is locally mapped to flipped AdS$_3$. Since the conifold is locally equivalent to $S^3$, its metric can be formally written as
\begin{equation}
    \begin{split}
        x^\mu & = l (\sin \chi \cos \psi, \cos \chi \cos \psi, \sin \psi \cos \xi, \sin \psi \sin \xi); \\
        \td s^2/l^2 & = \td \psi^2 + \cos^2 \psi \td \chi^2 + \sin^2 \psi \td \xi^2, ~~ \xi, \chi \in [0, 2 \pi),~ \psi \in [0, \pi/2],
    \end{split}
\end{equation}
but with the periodicity $(\xi, \chi) \simeq (\xi + 2 \pi \frac{r_+}{l}, \chi + 2 \pi i \frac{r_-}{l})$ given by \eqref{eq:Gamma+}, rather than the usual $2 \pi$ periodicity. The complexification $\psi \rightarrow i \rho$ is equivalent to $(x^3, x^4) \rightarrow i (z^3, z^4)$. In this case, the embedding \eqref{eq:S3embed} becomes
\begin{equation}
    \td s^2 = -(\td z^3)^2-(\td z^4)^2+(\td x^1)^2+(\td x^2)^2, \quad -(z^3)^2-(z^4)^2+(x^1)^2+(x^2)^2 = l^2.
\end{equation}
Therefore, after the complexification $\psi \rightarrow i \rho$, the conifold is locally equivalent to flipped AdS$_3$ space. Then, with the transformation
\begin{equation}
    \rho = \cosh^{-1} \sqrt{\frac{r^2+r_+^2}{r_+^2 + r_-^2}},
\end{equation}
one obtains
\begin{equation} \label{eq:AdSKerr}
    \td s^2 = - \frac{r^2 l^2 \td r^2}{(r^2 + r_+^2) (r^2 - r_-^2)} + l^2 \frac{r^2 + r_+^2}{r_+^2 + r_-^2} \td \chi^2 - l^2 \frac{r^2 - r_-^2}{r_+^2 + r_-^2} \td \xi^2,
\end{equation}
where $r_\pm$ are given in \eqref{eq:r+-}. Based on the periodicities of $(\xi, \chi)$ given in \eqref{eq:Gamma+}, we can introduce a complex linear transformation of the form
\begin{equation}
    \xi = \frac{r_+}{l} \phi + i \frac{r_-}{l^2} t, \qquad \chi = i \frac{r_-}{l} \phi + \frac{r_+}{l^2} t
\end{equation}
so that the periods of the new coordinate $\phi$ is normalized to $2 \pi$. Consequently, the metric becomes
\begin{equation} \label{eq:BTZlike}
    \td s^2 = - \frac{r^2 l^2 \td r^2}{(r^2 + r_+^2) (r^2 - r_-^2)} - r^2 \left( \td \phi - i \frac{r_+ r_-}{l r^2} \td t \right)^2 + \frac{(r^2 + r_+^2) (r^2 - r_-^2)}{r^2 l^2} \td t^2.
\end{equation}
This is the asymptotic flipped AdS$_3/\mathbb Z$ spacetime \eqref{asymflipped}.

The boundary of this metric is a Lorentzian torus defined at $r \rightarrow \infty$
\begin{equation}
    (\td s^2/r^2) |_{r \rightarrow \infty} = - \td \phi^2 + l^{-2} \td t^2 = \td z \td \bar z
\end{equation}
where the dual CFT$_2$ lives. Here we used the light-cone coordinates $(z, \bar z) = (\frac{t}{l} - \phi, \frac{t}{l} + \phi)$ to describe the left- and right-moving modes on the Lorentzian torus. The central charge of this dual CFT$_2$ is $c = \frac{3 i l}{2G}$, as the metric \eqref{eq:BTZlike} locally differs from AdS$_3$ by a Wick rotation $l\rightarrow -i l_{\text{AdS}}$.

The usual period $2 \pi$ of $(\xi, \chi)$ causes thermal effects in the asymptotic flipped AdS$_3/ \mathbb Z$ space, since both $\xi, \chi$ have imaginary parts. Note that $\xi, \chi$ can be expressed by the light-cone coordinates as
\begin{equation}
    \xi = \frac{r_+ + i r_-}{2l} \bar z - \frac{r_+ - i r_-}{2l} z, \qquad \chi = \frac{r_+ + i r_-}{2l} \bar z + \frac{r_+ - i r_-}{2l} z.
\end{equation}
The thermal periodicity in the asymptotic fAdS$_3$ corresponding to that in Kerr-dS$_3$ \eqref{eq:thermalchi} is $(\xi, \chi) \simeq (\xi,\chi + 2\pi)$, which is equivalent to
\begin{equation}
    z \simeq z + \frac{2 \pi l}{r_+ - i r_-}, \qquad \bar z \simeq \bar z + \frac{2 \pi l}{r_+ + i r_-}.
\end{equation}
The corresponding modular parameters are therefore
\begin{equation}
    \tau = \frac{1}{-2 \pi} \frac{2 \pi l}{r_+ - i r_-} = -\frac{l}{r_+ - ir_-} \,, \qquad \bar{\tau} = \frac{1}{2 \pi} \frac{2 \pi l}{r_+ + i r_-} = \frac{l}{r_+ + ir_-}.
\end{equation}
which are exactly equivalent to the modular parameters obtained by the alternative method in Eq.~\eqref{eq:modular}. Consequently, the Cardy formula also yields the same entropy $S = \frac{2 \pi r_+}{4G}$.

\bibliographystyle{JHEP}
\bibliography{biblio.bib}
\end{document}